\documentclass[aps,reprint,twocolumn,pra,groupedaddress]{revtex4-1}
\usepackage{amsmath,amssymb,amsfonts}
\usepackage{mathrsfs}
\usepackage{hyperref}
\usepackage{soul}
\usepackage{graphicx}
\usepackage{color}
\begin{document}
\title{Nonlinear nonuniform $\mathcal{PT}$-symmetric  Bragg grating structures}
\author{S. Vignesh Raja}
\email{vickyneeshraja@gmail.com}
\author{A. Govindarajan}
\email{govin.nld@gmail.com}
\author{A. Mahalingam}
\email{drmaha@annauniv.edu}
\author{M. Lakshmanan}
\email{lakshman.cnld@gmail.com}
\affiliation{$^{*,\ddagger}$Department of Physics, Anna University, Chennai - 600 025, India}
\affiliation{$^{\dagger,\mathsection}$Centre for Nonlinear Dynamics, School of Physics, Bharathidasan University, Tiruchirappalli - 620 024, India}
\begin{abstract}
We explore the consequences of incorporating parity and time reversal ($\mathcal{PT}$) symmetries on the dynamics of nonreciprocal light propagation  exhibited by a class of nonuniform periodic structures known as chirped $\mathcal{PT}$-symmetric fiber Bragg gratings (FBGs). The interplay among various grating parameters such as chirping, detuning, nonlinearities, and gain/loss gives rise to unique bi- and multi-stable states in the unbroken as well as broken $\mathcal{PT}$-symmetric regimes. The role of chirping  on the steering dynamics of the hysteresis curve is influenced by the type of nonlinearities and the nature of  detuning parameter. Also, incident directions of the input light robustly impact the steering dynamics of bistable and multistable states both in the unbroken and broken $\mathcal{PT}$-symmetric regimes. When the light launching direction is reversed, critical stable states are found to occur at very low intensities which opens up a new avenue for an additional way of controlling light with light. We also analyze the phenomenon of unidirectional wave transport and the reflective bi- and multi-stable characteristics at the so-called $\mathcal{PT}$-symmetry breaking point.
\end{abstract}
\maketitle
\section{Introduction} 
Fiber Bragg gratings are regarded as one of the indispensable components in all-optical communication systems and sensor networks, see \cite{hill1997fiber, erdogan1997fiber,othonos1997fiber,giles1997lightwave} and references therein.
Numerous theoretical and experimental investigations were carried out in the past, in uniform \cite{winful1979theory} as well as nonuniform grating structures \cite{radic1995theory,radic1994optical, maywar1998effect}   to build efficient switches \cite{ping2005nonlinear,zang2012}, regenarators \cite{8687563}, memory devices \cite{karimi2012all} and so on \cite{ping2005bistability, yosia2007double}, based on the phenomena of optical bistability (OB) and multistability (OM).   

Optical bistability is a distinctive attribute of nonlinear periodic structures with a distributive feedback mechanism \cite{winful1979theory}. The existence of intensity dependent refractive index alone is not adequate to manifest OB \cite{gibbs2012optical} in systems like couplers \cite{jensen,govindaraji2014dark,govindaraji2015numerical}. The feedback provided by the system in combination with the nonlinearity of the device facilitates the output of the systems such as FBGs \cite{yosia2007double, broderick1998bistable}, resonators \cite{rukhlenko2010analytical}, and anti-directional couplers \cite{litchinitser2007optical, govindarjan2019} to be bi (multi) stable.  FBGs exhibit an inherent stop band which does not allow certain wavelengths to get transmitted \cite{broderick1998bistable}. This stop band can be easily tuned by launching an external signal in the form of laser \cite{karimi2012all} or a soliton pulse \cite{yousefi2014all} into it. When the incident pulse intensity increases, the reflection (transmission) manifests either as blue shift or as redshift from its resonance location due to the intensity dependence of the refractive index \cite{winful1979theory}. Simultaneously, there is an increase in the internal optical intensity due to the detuning dependent feedback mechanism \cite{maywar1998low}.  Such a build up in the intensity shifts the spectrum further. For a finite value of detuning, there is a sudden jump in the output intensity indicating an up-switching action. When the input intensity decreases, the converse effect occurs, but at a different input intensity, as the optical intensity inside the device is adequate to sustain the input off-tuned from its original location \cite{karimi2012all, maywar1998effect}. Hence, it results in the formation of a hysteresis curve in the output characteristics of the system. 

When it comes to all-optical storage applications, the width of the hysteresis curve and range of wavelengths over which OB and OM curves occur (spectral span) should be considerable \cite{maywar1997transfer}.  If the OB curves meet these requirements, the input power at which the stable state appears can serve as the control pulse to switch between the \emph{on} and \emph{off} states of flip-flops \cite{karimi2012all}. In addition to these, chirped nonlinear FBGs are well known to offer a functionality known as spectral uniformity \cite{maywar1998low}. A OB/OM system is said to be offering large spectral uniformity if it satisfies the following conditions: There should be a range of input powers, which are common to different spectral components (variation in the detuning parameter $\delta$) before switching and their corresponding \emph{on} and \emph{off} states should be separated by a significant amount of input powers \cite{maywar1997transfer, maywar1998low,maywar1998effect}.  Physically, chirping denotes the variation in the grating period ($\Lambda$) as a function of position ($z$). In other words, the Bragg wavelength of the device is not constant any longer, but it gets altered gradually along the longitudinal distance ($z$) of the FBG \cite{hill1994chirped, erdogan1997fiber}. It is worthwhile to mention that for switching applications, the chirping brings in notable variations in the intensities (up and down) at which the switching takes place \cite{kim2001effect}.

The advancements in the current optical technologies fetch one more degree of freedom to engineer the nature of light propagation in an optical system by taking advantage of the interaction among the linear tunneling, intrinsic loss and the extrinsic gain of the system \cite{ozdemir2019parity, el2018non, longhi2018parity,feng2016}.  Such an interplay is feasible only if the refractive index of the medium is taken to be a complex entity rather than real \cite{el2007theory}. In particular, invoking the notion of $\mathcal{PT}$-symmetry in any  conventional optical systems  is itself a new physics as it leads to novel non-Hermitian systems which are supposed to be a trending topic in optics. The pioneering works on the $\mathcal{PT}$-symmetric notion are devoted to the exploration on the dynamics of systems in different $\mathcal{PT}$-symmetric regimes (see \cite{bender1998real, kottos2010optical, ruter2010observation,lin2011unidirectional,sukh2010,kominis2017,borispt2019,boris2013} and references therein). Quite recently, a work on  $\mathcal{PT}$-symmetric couplers reveals the potential influence of $\mathcal{PT}$-symmetry in the reduction of switching intensities and unwanted steering oscillations \cite{govindarajan2018tailoring}. In a linear FBG, delicate balance between gain and loss induces nonreciprocal reflection response \cite{kulishov2005nonreciprocal}, unidirectional invisibility \cite{lin2011unidirectional}, coherent perfect absorption and lasing (CPAL) \cite{baum2015parity, hahn2016unidirectional}. Not long ago, Lupu \emph{et al.} demonstrated the frequency discriminative properties of an apodized $\mathcal{PT}$-symmetry FBG \cite{lupu2016tailoring}.  As of now, investigations on optical bistability manifested by a $\mathcal{PT}$-symmetric FBG confines to the nonlinear uniform gratings with third-order nonlinearity alone \cite{1555-6611-25-1-015102}, saturable and dispersive gain/loss profiles \cite{phang2014impact}, and modulated Kerr nonlinearity \cite{komissarova2019pt}. To date, there seems to exist no works in the literature dealing with the OB in chirped $\mathcal{PT}$-symmetric FBGs \cite{komissarova2019pt, vigneshraja2019}. Also, there exist a few articles which illustrate the direction dependent OB phenomenon \cite{komissarova2019pt}. Motivated by these facts, we investigate OB/OM for right and left incidences of light in a $\mathcal{PT}$-symmetric chirped FBG in order to highlight potential applications in all-optical switching and memories. In particular, we took a step ahead to establish an interplay between the chirping and detuning parameters in such a nonuniform $\mathcal{PT}$-symmetric FBG which has not so far been dealt with in the literature (as far as our knowledge goes). It is believed that any practical system of interest operating in the broken $\mathcal{PT}$-symmetric regime may encounter instabilities \cite{komissarova2019pt}. But it is shown in the present work that the novel  optical bistable and multistable phenomena persist even in the presence of nonuniformites of the grating with no signs of instability in the broken $\mathcal{PT}$-symmetry regime.

To do so, we structure the article as follows. Section \ref{Sec:2} describes the mathematical model of the system. Sections \ref{Sec:3} explores the bi and multistability features offered by the system in the unbroken  and broken $\mathcal{PT}$-symmetric regimes for different directions of light incidence under various nonlinear regimes. Section \ref{Sec:4} elucidates the exceptional point dynamics of the system with a special mention to reflective OB/OM. The conclusion is drawn in Sec. \ref{Sec:5}.

\section{Model}\label{Sec:2}
To start with, we consider a non-uniform FBG with length ($L$) and grating period ($\Lambda$) inscribed into a chirped chalcogenide fiber. The complex refractive index distribution profile that  describes such a system is mathematically written as \cite{PhysRevA.86.033801, yousefi2015all,erdogan1997fiber} 

\begin{gather}
\nonumber n(z)=n_{0}+n_{1R}\cos\left(\frac{2\pi}{\Lambda}z+\phi(z)\right)\\+in_{1I}\sin\left(\frac{2\pi}{\Lambda}z+\phi(z)\right)+n_{2}|E|^{2}+n_{4}|E|^{4}
\label{Eq:Norm1}
\end{gather}

Here, $n_0$ stands for the refractive index of the core and $n_{1R}$ describes the strength of modulation of the grating. The gain/loss is included into the system through the factor $n_{1I}$ and 
$\phi(z)$ describes the grating phase with respect to longitudinal distance ($z$) \cite{erdogan1997fiber}. Also,  $n_2$ and $n_4$ indicate the cubic and quintic nonlinear coefficients, respectively. The transverse electric field $E(z)$ inside the FBG is the superposition of two counter propagating modulated modes which reads as \cite{PhysRevA.86.033801}
\begin{gather} 
E (z)= E_{f}(z) \exp \left[i\beta_{0} z\right]  +E_{b}(z) \exp \left[-i\beta_{0} z\right] 
\label{Eq:Norm2}
\end{gather}
where $E_f$ and $E_b$ are the slowly varying amplitudes of the forward and backward electric fields, respectively. The propagation constant is given by $\beta_{0} = 2 \pi n_{0} / \lambda_{0}$, where  $\lambda_{0}$ is the operating wavelength.

The coupled mode equations that describe the proposed system in normalized form are written as \cite{sarma2014modulation,komissarova2019pt}
\begin{gather}
+i\frac{d \psi_{+}}{d\zeta}+{\overline{\delta}}\psi_{+}+\left({k}\pm{g}\right)\psi_{-}+{\gamma}\left(|\psi_{+}|^{2}+2|\psi_{-}|^{2}\right)\psi_{+}-\nonumber\\{\varGamma}\left(|\psi_{+}|^{4}+6|\psi_{+}|^{2}|\psi_{-}|^{2}+3|\psi_{-}|^{4}\right)\psi_{+}
=0, \label{Eq:norm7}\\
-i\frac{d \psi_{-}}{d\zeta}+{\overline{\delta}}\psi_{-}+\left({k}\mp{g}\right)\psi_{+}+{\gamma}\left(|\psi_{-}|^{2}+2|\psi_{+}|^{2}\right)\psi_{-}-\nonumber\\{\varGamma}\left(|\psi_{-}|^{4}+6|\psi_{+}|^{2}|\psi_{-}|^{2}+3|\psi_{+}|^{4}\right)\psi_{-}
=0. \label{Eq:norm8}
\end{gather}
where $\psi_{\pm}$ and $\zeta$ are the normalized forward, backward field components, and spatial coordinate respectively. The terms $k, g, \gamma, \Gamma$, and $\delta $ accordingly represent the coupling, gain/loss, cubic nonlinearity, quintic nonlinearity, and detuning parameters \cite{PhysRevA.86.033801}. The sign $+$ ($-$) in the third term of Eq. \eqref{Eq:norm7} indicates the launching directions of light on the left (right) input surface of the FBG. It is to be noted that if $k>g$, the system is said to be operated in the unbroken $\mathcal{PT}$-symmetric regime. On the other hand, if $g>k$, the system is set to be working in the broken $\mathcal{PT}$-symmetric regime. The mathematical condition in which both the coupling and gain/loss coefficients are equal ($k=g$) is termed as  the \emph{exceptional point} or \emph{singularity}. Also, the detuning parameter is not a constant as in the case of chirped FBGs  and it reads as \cite{maywar1998effect}
\begin{gather}
 \overline \delta (\zeta) = \delta + C {(\zeta-L / 2)}/{L^{2}}.
 \label{Eq:norm9}
 \end{gather}
Note that the normalized parameter $\left(\overline{\delta}\right)$ is a function of spatial coordinate ($\zeta$) and it includes the effect of normalized spatial chirping (C) as well as the normalized detuning parameter ($\delta$). The coupled mode equations (\ref{Eq:norm7}) and (\ref{Eq:norm8}) are solved numerically by implicit Runge-Kutta fourth order method with the following initial conditions
\begin{gather}
\psi_{+}(0)=\psi_{0}, \qquad \psi_{-}(L)=0.
\end{gather}

 Throughout the paper we use the system parameters as $L = 2$ and $ k =4$ unless specified. In the cubic nonlinear regime, we have $\gamma = 1$ and $\Gamma = 0$, while in the quintic nonlinear regime, the coefficients are assumed to be, $\gamma = \Gamma = 1$. The  intensity of the input laser pulse is given by $P_0 = |\psi_+(0)|^2$. Similarly, the output and transmitted intensities of the system are noted by the relations, $P_1(L) = |\psi_{+}(L)|^2$ and $P_1(L)/P_0$ = $|\psi_+(L)/\psi_+(0)|^2$, respectively.

\section{Results and discussions}
\label{Sec:3}
\begin{figure}[t]
	\centering
	\includegraphics[width=0.5\linewidth]{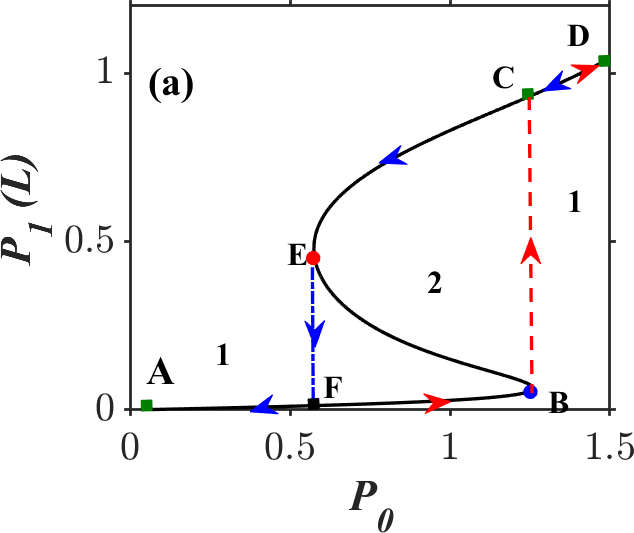}\includegraphics[width=0.49\linewidth]{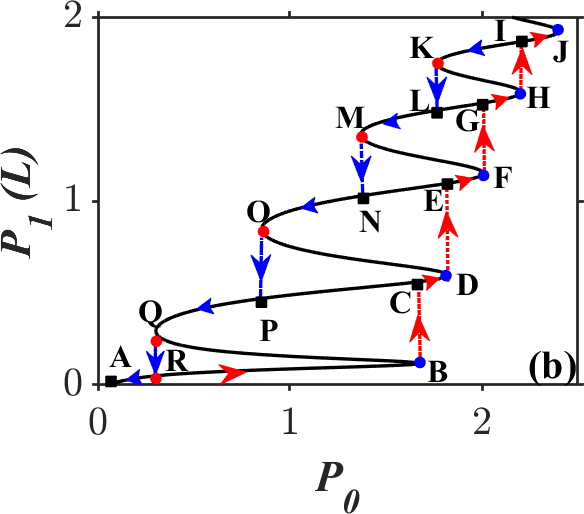}\\
	\caption{Schematic sketches showing different stable/unstable regions of typical (a) OB and (b) OM curves in a chirped $\mathcal{PT}$-symmetric FBG. 
	} 
	\label{fig0}    
\end{figure}
\begin{figure}[t]
	\centering
\includegraphics[width=0.5\linewidth]{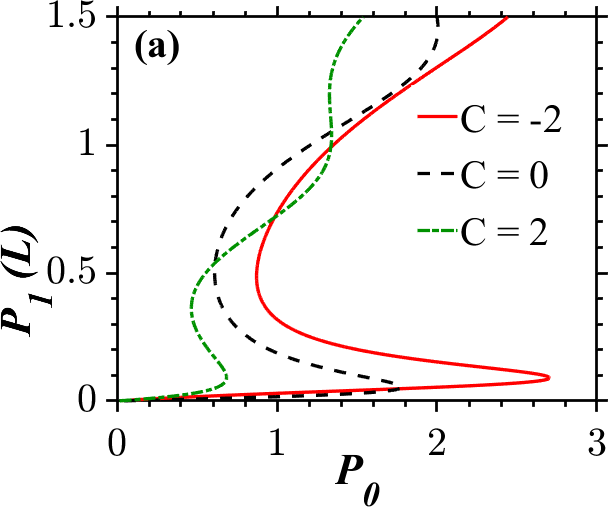}\includegraphics[width=0.5\linewidth]{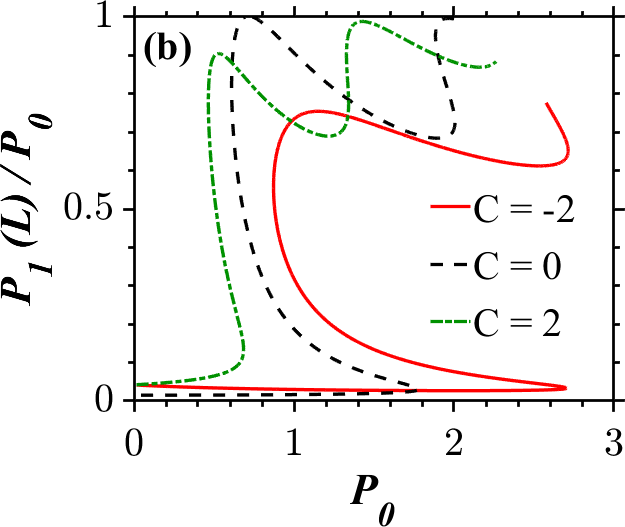}\\\includegraphics[width=0.5\linewidth]{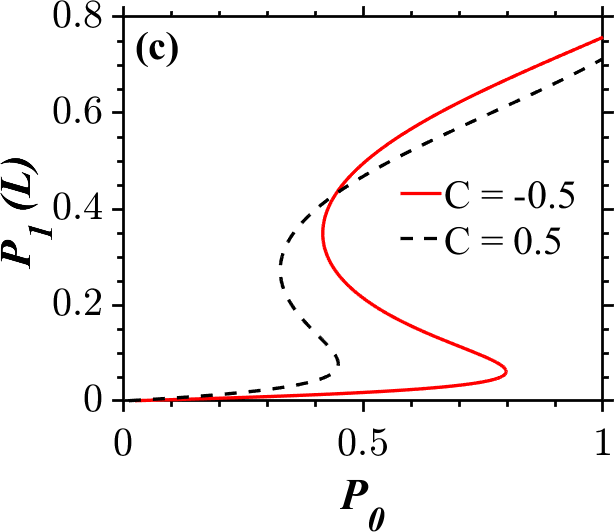}\includegraphics[width=0.5\linewidth]{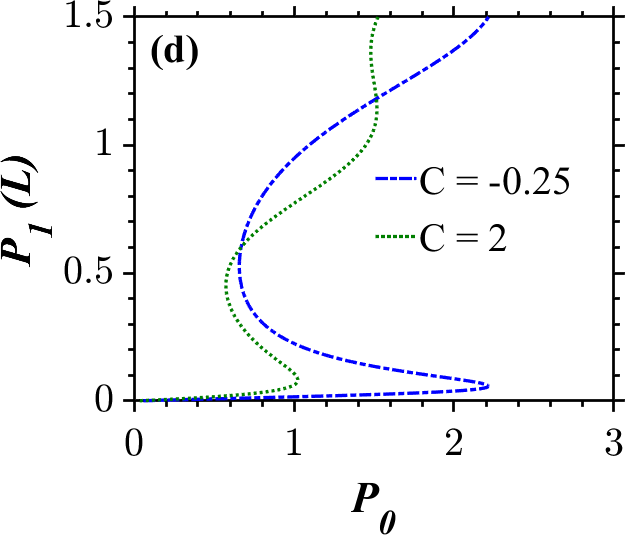}\\\includegraphics[width=0.5\linewidth]{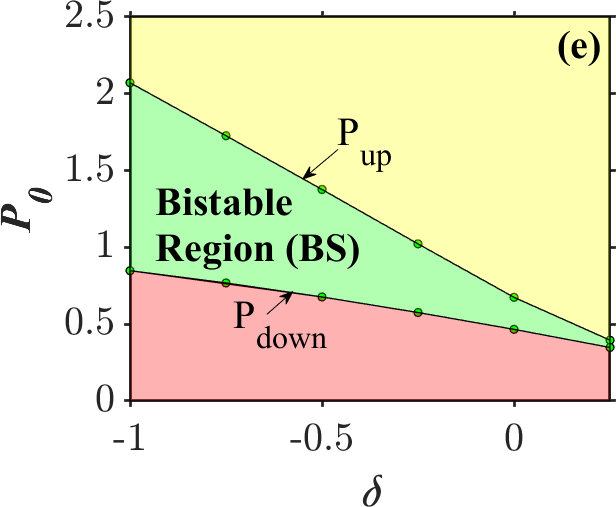}\includegraphics[width=0.5\linewidth]{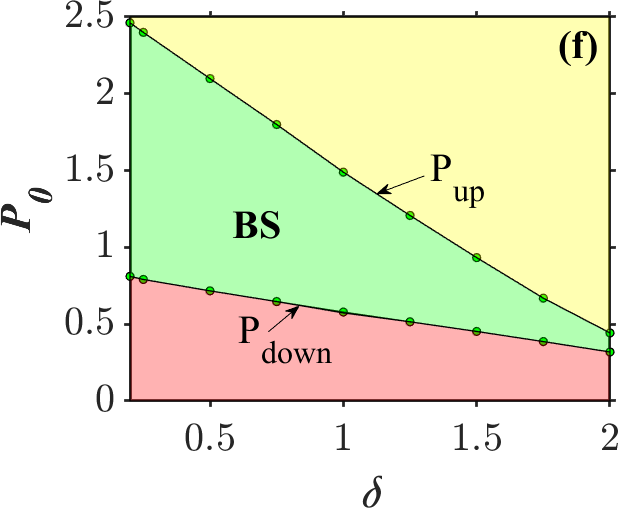}
	\caption{Plots showing (a) OB in a chirped $\mathcal{PT}$-symmetric FBG in the unbroken regime ($g =3.75$) for the left incidence in the presence of cubic nonlinearity. The curves are simulated at  $\delta = 0$ for different  chirping (C) values and (b) depicts the transmission curves corresponding to (a). The interplay between chirping and  detuning parameter ($\delta = 1$ and $-0.25$) is shown in plots (c) and (d), while plots (e) and (f) illustrate the variation in switching intensities against detuning parameter ($\delta$) at $C = 2$ and $-2$, respectively. The shaded regions in red and yellow indicate $P_0 < P_{down}$ and $P_0 > P_{up}$, respectively. The values of $P_0$ for which the system shows the bistable behavior are shaded in green.} 
	\label{fig_1}    
\end{figure}

\begin{figure}[t]
	\centering
	\includegraphics[width=0.5\linewidth]{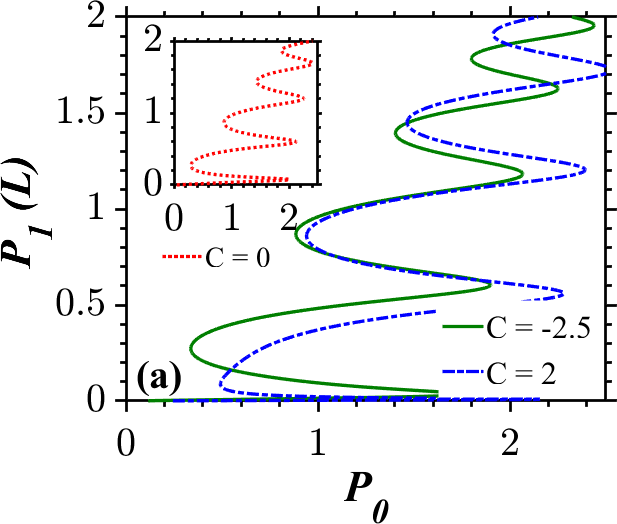}\includegraphics[width=0.5\linewidth]{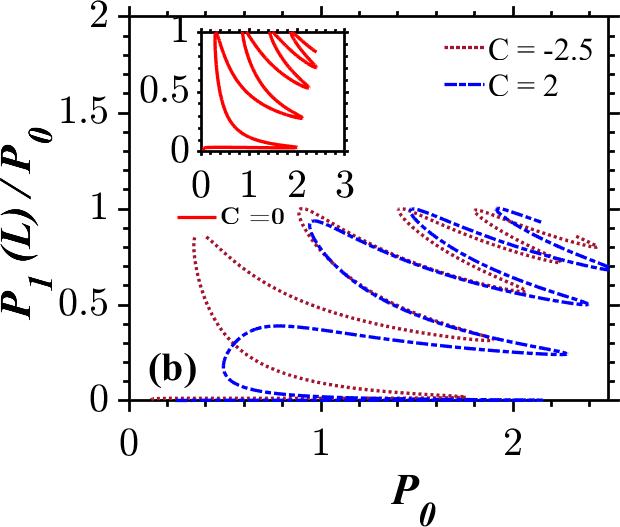}\\\includegraphics[width=0.5\linewidth]{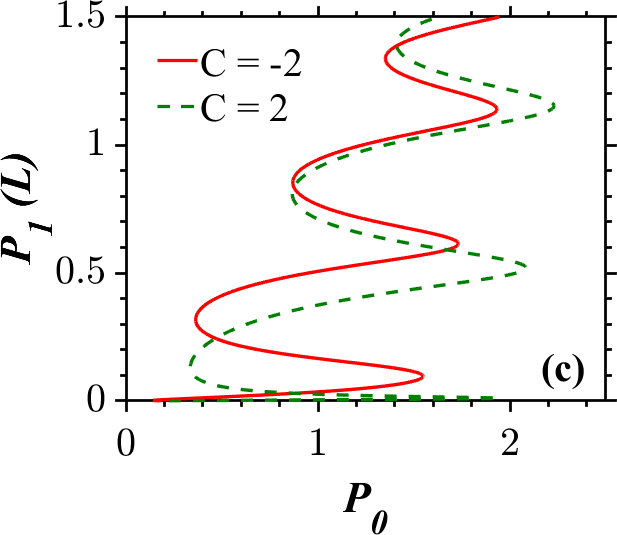}\includegraphics[width=0.5\linewidth]{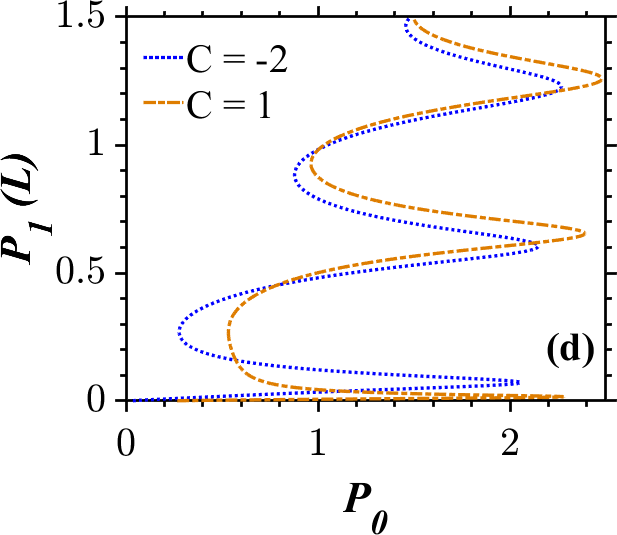}\\\includegraphics[width=0.5\linewidth]{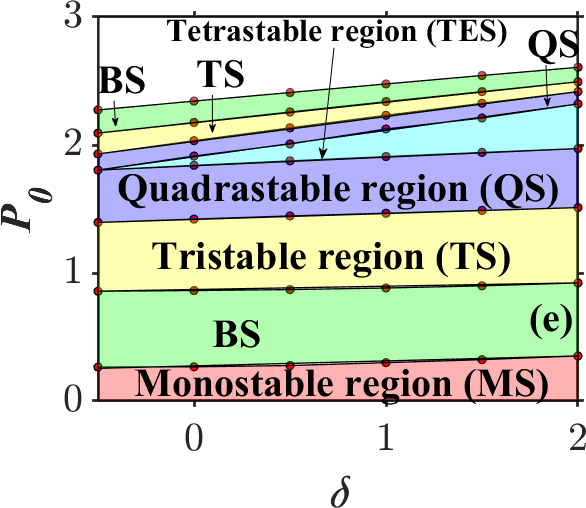}\includegraphics[width=0.5\linewidth]{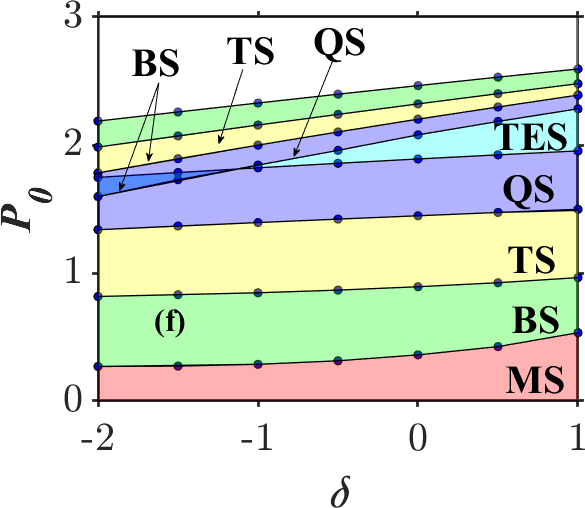}
	\caption{OM in a chirped $\mathcal{PT}$-symmetric FBG in the unbroken regime ($g = 2$) for the left incidence in the presence quintic nonlinearity. Plots (a) and (b) portray the role of chirping (C) on the input-output and transmission characteristics of the system, respectively at $\delta = 0$. The variation in the OM curves for different values of chirping (C) at finite detuning parameter ($\delta = -1$ and $1$) is shown in plots (c) and (d) whereas plots (e) and (f) illustrate the different stable regions in the $\delta$ vs $P_0$ curves at $C = -1$ and $1$, respectively. The red shaded region indicates single stable state while the regions shaded in green, yellow, blue, cyan represent the range of input powers ($P_0$) for which the system possesses two, three, four and five stable states, respectively. }
	\label{fig2}    
\end{figure}
To explain the phenomenon of OB and OM, typical OB and OM curves are shown in Figs. \ref{fig0}(a) and \ref{fig0}(b), respectively. In Fig. \ref{fig0}(a) AB and DE are the stable branches whereas BE is an unstable branch. When the input intensity is increased from zero, the system remains in the first stable branch till the point B and above which the system switches to the second branch (CD) with any small increase in the value of $P_0$.  When the input intensity is decreased, the system remains in the stable branch DE. If $P_0$ is reduced below E, the system switches back to point F in the first stable branch. In the region FBCE, we observe that for any input intensity there exist two output intensities and this region is said to be the bistable region. Similarly, in Fig. \ref{fig0}(b) AB, DQ, FO, HM, JK are
the stable branches and BQ, DO, FM, KH indicate the
unstable branches.

\subsection{Left incidence: Unbroken $\mathcal{PT}$-symmetric regime}
The chirping parameter (C) serves as an additional degree of freedom to significantly alter the feedback mechanism that is responsible for the OB/OM phenomenon. Such OB/OM variations resulting in variations in intensity, size and shape of the hysteresis loop arise from the dependence of internal feedback mechanism of the system on the Bragg wavelength. To understand the role of chirping in a $\mathcal{PT}$-symmetric FBG, we first analyze the OB in the absence of detuning parameter ($\delta = 0$) in the unbroken $\mathcal{PT}$-symmetric regime for the left light incidence. We infer from Fig. \ref{fig_1}(a) that the negative chirping parameter broadens the hysteresis curves and increases the switching intensities. On the other hand, for positive values of the chirping parameter, the intensities to switch between the bistable states decrease and so we observe a narrow hysteresis curve. The physical explanation for such a phenomenon is straight forward to figure out. In the presence of self-focusing nonlinearity, the positive chirp enhances the feedback of the system, while the negative chirp plays an adverse role of weakening the feedback of the system.  Also, in the absence of chirping, the transmission peak is unity. The chirping parameter reduces the amount of light transmitted by the $\mathcal{PT}$-symmetric device for positive chirping values as seen in Fig. \ref{fig_1}(b).  The same effect is observed for negative chirping parameters too. In particular, the chirping drastically reduces the amount of transmitted light  when $C = -2$. However, for the same input power any increase in the value of positive chirping ($C = 2$) results in the formation of multiple transmission peaks. It is important to note that the detuning parameter, which is regarded as the inevitable control parameter in linear FBG system \cite{hill1997fiber,erdogan1997fiber,othonos1997fiber,giles1997lightwave},  is often overlooked in many scientific contributions dealing with nonlinear switching phenomenon as they are restricted to investigate the switching phenomenon at the Bragg wavelength \cite{lin2011unidirectional,komissarova2019pt} or wavelengths very closer to it (order of 0.1 nm) \cite{yousefi2015all,yosia2007double,ping2005bistability,karimi2012all}.  The reason for selection of such closer wavelength lies in the fact that no optical source (including lasers) is perfectly monochromatic but they posses a narrow line width. As a consequence, operating at the synchronous wavelength (or near) may impose an additional problem to have a control mechanism  in order to maintain ideal phase matching. By definition, the detuning refers to the condition where the operating wavelength of the grating lies away from the Bragg wavelength and such detuned systems are supposed to be physically more  interesting and realistic than  ideal systems. Indeed it has been recently reported that the detuning parameter in the Manakov model supports novel non-degenerate solitons \cite{stalin2019}.  Also, as in the case of conventional FBG, the role of the former has been clearly pointed out that in the absence of higher order nonlinearities, the positive detuning parameter lessens the switching intensities, whereas the negative detuning parameter increases them \cite{zang2012}. 

These consequences hold true even for the $\mathcal{PT}$-symmetric chirped FBG as seen in Figs. \ref{fig_1}(c) and \ref{fig_1}(d). Note that physically, the operating wavelength of the grating should lie 10 nm below or above the Bragg wavelength so as to exhibit OB curves. Otherwise if the incoming field wavelength were to fall outside the stopband of the grating, the feedback mechanism will be insufficient to induce OB phenomenon \cite{kim2001effect,zang2012,broderick1998bistable}.  Till now, we have discussed the individual effect of chirping and detuning in the absence of the other.  Since both these parameters play significant roles in controlling the feedback mechanism,  it is obvious that reduction or enhancement in the switching intensity will bound to happen inside a chirped $\mathcal{PT}$-symmetric FBG.  As detuning and chirping parameters can have both positive and negative values,   we now look into their collective effect for four different scenarios. When both $C$ and $\delta$ have positive signs, an OB curve with very low switch-up intensity and minimal width is observed (black dotted OB in Fig. \ref{fig_1}(c)).  However, when both these parameters have negative signs, the hysteresis curve is much wider and the switching intensities are also high (brown dotted OB in Fig. \ref{fig_1}(d)). For the other two cases (green solid and red solid in the middle panel), featuring $C$ and $\delta$ with opposite signs, the values of the switching intensities lie between the intensities offered by the system with the same signs of $C$ and $\delta$. The positive value of chirping results in the formation of desirable bistable states in a narrow spectral range with low switching intensities as shown in Fig. \ref{fig_1}(e). But the range of wavelengths which permit the low intensity switching is much broader than that of the system simulated in the absence of chirping (not shown here).  This is feasible because the wavelengths which featured low feedback in the absence of chirping are now subjected to strong feedback as a consequence of variation in the Bragg wavelength. The spectral range broadens besides an increase in the switching intensities when the chirping parameter is negative as in the case of Fig. \ref{fig_1}(f).

With the addition of quintic nonlinearity, the system exhibits multistable states in its input-output characteristics as shown in Fig. \ref{fig2}(a). When compared to the operation in the absence of chirping ($C = 0$), there is a marginal decrement and increment in the switching intensities and width of the curves corresponding to the negative ($C = -2.5$) and positive chirping ($C = 2$) values. One can also observe from Fig. \ref{fig2}(b) that the transmission at the first peak reduces from unity to lower values when the chirping parameter ($C$) takes some finite values. 
We reason out that due to the presence of self-defocusing quintic nonlinearity, the system exhibits switching phenomenon at low intensities when both $C$ and $\delta$ take negative values (red solid) as seen in Fig. \ref{fig2}(c). For the same value of detuning, the switching occurs at higher intensities if the chirping is positive (green dashed). Out of the four curves in the bottom panel of Fig. \ref{fig2}, the highest values of switching intensities are observed for the curve (brown solid) whose chirping value is positive besides operating in the longer wavelength side as shown in Fig. \ref{fig2}(d). Compared to this curve, the multistable curve plotted with a negative value of chirping and positive detuning (blue dashed) features marginally low switching intensities. When the input power is tuned from zero to a finite value (say $P_0 = 2.5$), the system shows different stable regions ranging from bistable to tetrastable as depicted in Figs. \ref{fig2}(e), \ref{fig2}(f) and Table \ref{tab3}. The positive chirping supports more number of stable states in the shorter wavelength side, while the negative detuning favors the same in the longer side of Bragg wavelength ($\delta=0$). With the ever-growing demand in the market to expand the transmission capacity of the network, multi-level signal processing schemes are inevitable in the next generation all-optical networks. Even though the multi-level signals can theoretically handle more data at a given symbol rate, their physical realization is hindered by the fact that the signal to noise ratio will degrade and bit error rate is likely to increase if the power level separating different signaling states are not distinct. In such a scenario, multistable states with good degrees of spectral uniformity emanating from the proposed system or a similar configuration can be used to set up all-optical regenerators to improve the quality of degraded signal.

\begin{table}[t]
	\caption{\centering{Number of stable states vs $P_0$ in the unbroken $\mathcal{PT}$-symmetric regime as observed in Figs. \ref{fig2}(e) and \ref{fig2}(f).}}
	\begin{center}
\begin{tabular}{|c|c|c|}
\hline
	{Number of}& {Region 1}& {Region 2} \\
{stable states}&{}&{} \\
\cline{2-3} 	
\hline
		 
{2}&$P_{down}^1$ $\le$ $P_0$ $<$ $P_{down}^2$& $P_{up}^3$ $<$ $P_0$ $\le$ $P_{up}^4$   \\
\hline
 {3}&$P_{down}^2$ $\le$ $P_0$ $<$ $P_{down}^3$& $P_{up}^2$ $<$ $P_0$ $\le$ $P_{up}^3$   \\
 \cline{2-3}
 {}&\multicolumn{2}{|c|}{$P_{up}^1$ $<$ $P_0$ $<$ $P_{down}^4$ (only for Fig. \ref{fig2}(e))}\\

\hline
	 {4}&$P_{down}^3$ $\le$ $P_0$ $<$ $P_{down}^4$& $P_{up}^1$ $<$ $P_0$ $\le$ $P_{up}^2$  \\
	\hline
	 {5}&\multicolumn{2}{|c|}{$P_{down}^4$ $\le$ $P_0$ $\le$ $P_{up}^1$} \\
\hline
\end{tabular}
\label{tab3}
\end{center}
\end{table}

\subsection{Left incidence: Broken $\mathcal{PT}$-symmetric regime}
\begin{figure}[t]
	\centering
	\includegraphics[width=0.5\linewidth]{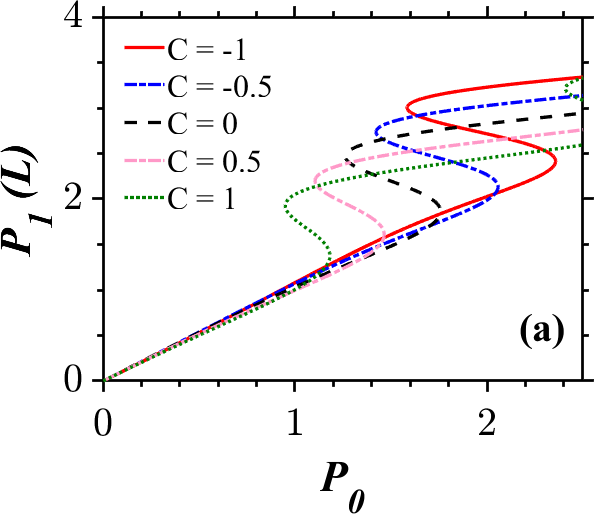}\includegraphics[width=0.5\linewidth]{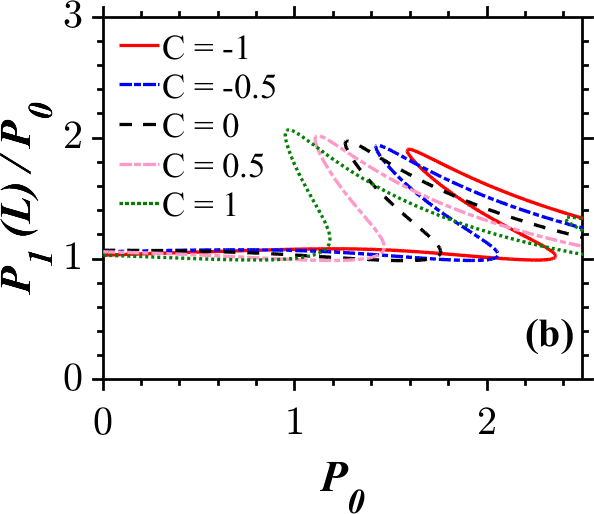}\\\includegraphics[width=0.5\linewidth]{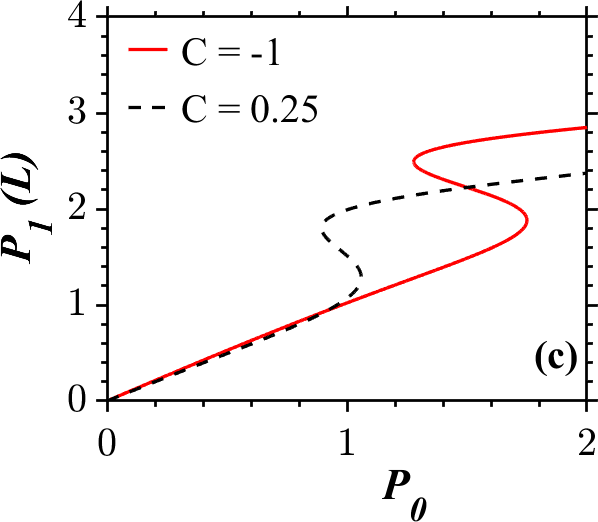}\includegraphics[width=0.5\linewidth]{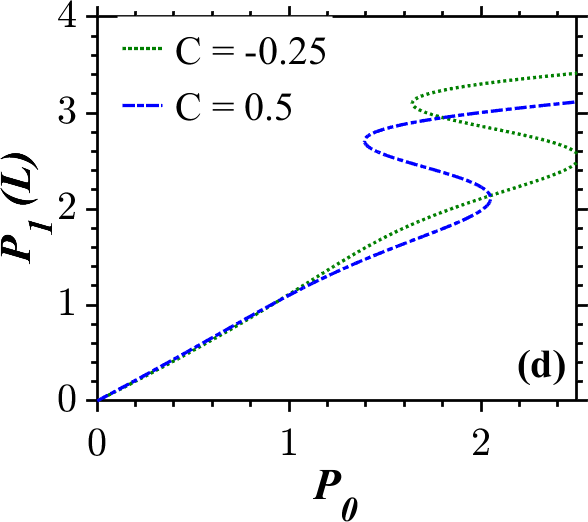}\\\includegraphics[width=0.5\linewidth]{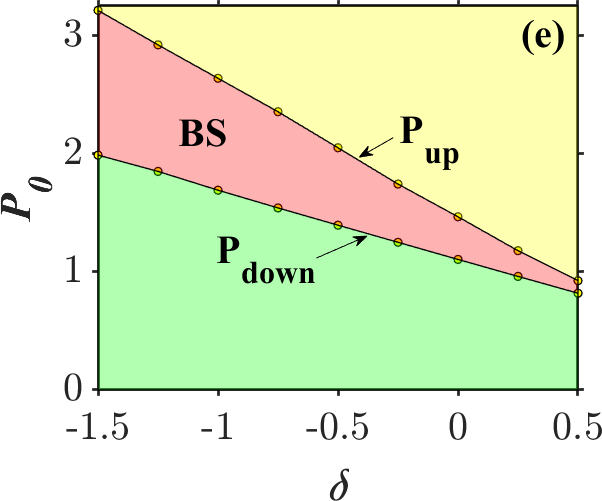}\includegraphics[width=0.5\linewidth]{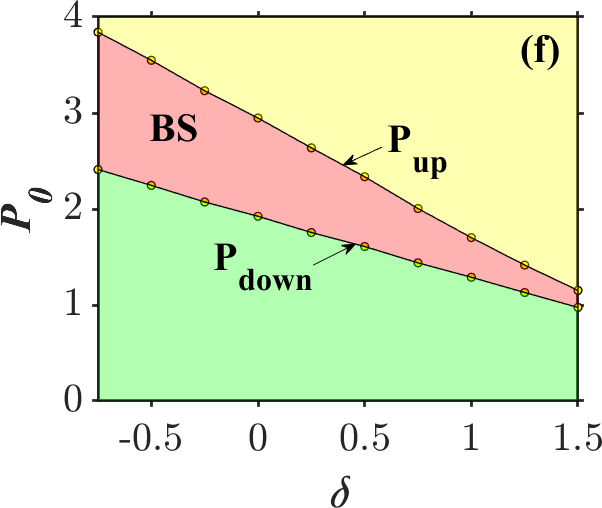}
	\caption{Plots corresponding to the novel bistable dynamics of the chirped FBG in the broken $\mathcal{PT}$-symmetric regime ($g=5$) with the same simulation parameters as in Fig. \ref{fig_1} except the following: The value of detuning parameter is $\delta = 0.5$ and $-0.5$ in plots (c) and (d), respectively and the value of chirping parameter is $C = 0.5$ in plot (e). In plots (e) and (f), the regions shaded in green and yellow indicate $P_0 < P_{down}$ and $P_0 > P_{up}$, respectively and the region shaded in red color represents the value of $P_0$ for which the system shows the bistable behavior.}
	\label{fig3}    
\end{figure}

\begin{figure}[t]
	\centering
	\includegraphics[width=0.5\linewidth]{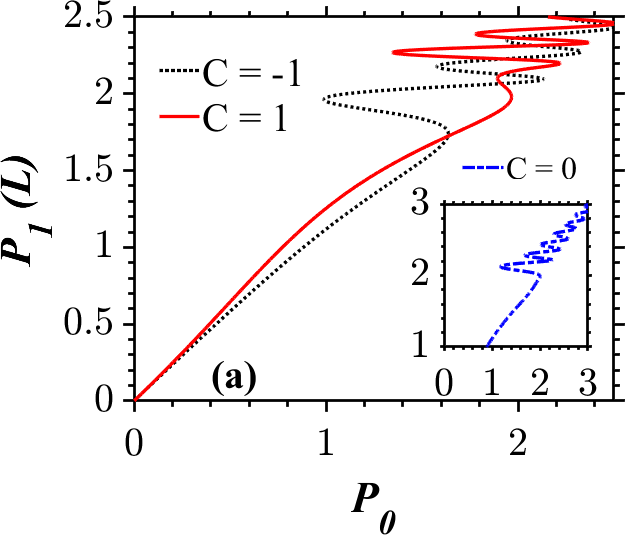}\includegraphics[width=0.5\linewidth]{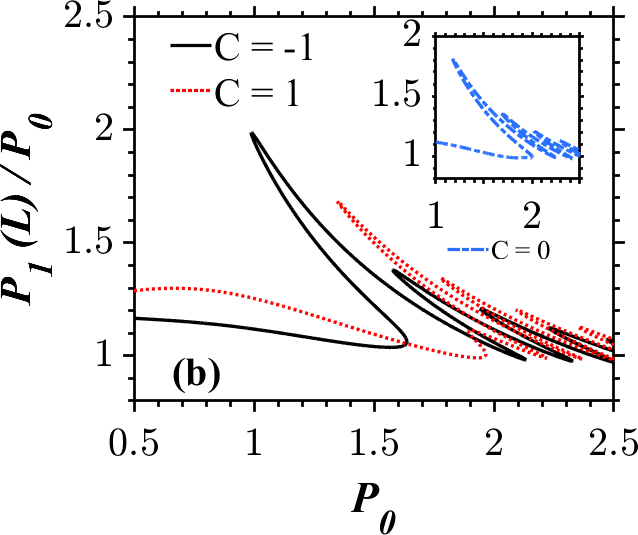}\\\includegraphics[width=0.5\linewidth]{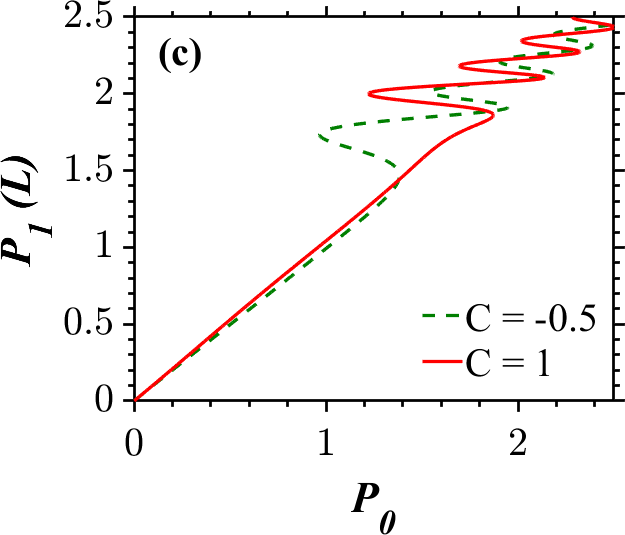}\includegraphics[width=0.5\linewidth]{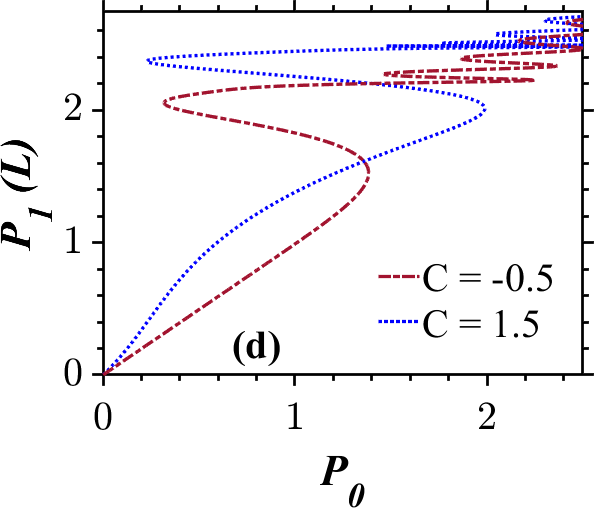}\\\includegraphics[width=0.5\linewidth]{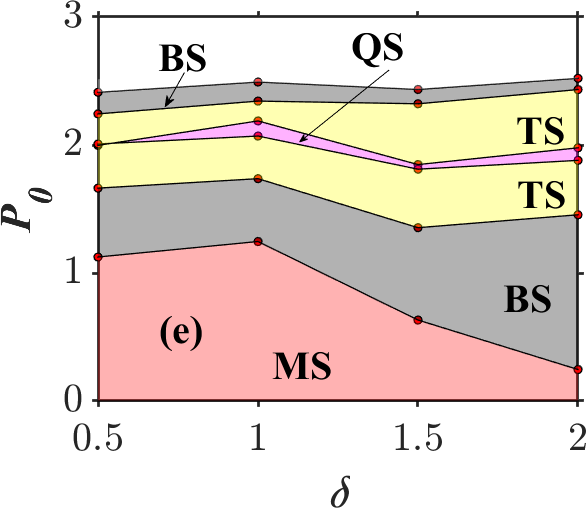}\includegraphics[width=0.5\linewidth]{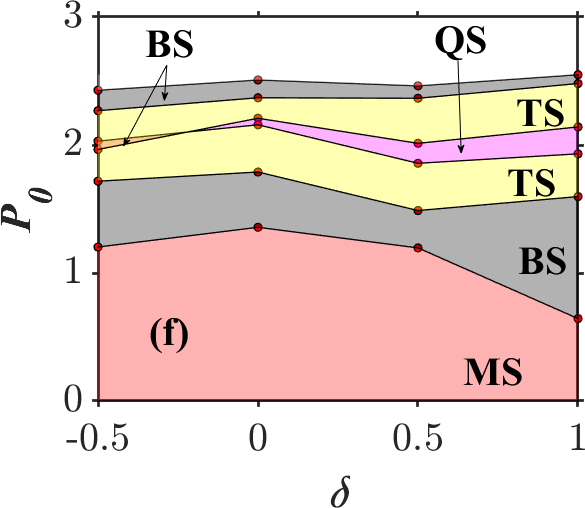}.
	\caption{Plots (a) and (b) show the same dynamics with the same system parameters used in Fig. \ref{fig2} in the broken $\mathcal{PT}$-symmetric regime ($g=5.2$).  In plots (e) and (f), the regions shaded in red refer single stable state and the regions shaded in black, yellow, magenta represent the range of input intensities ($P_0$) over which the system possesses two, three and four output stable states, respectively.
}
	\label{fig4}    
\end{figure}

As of now, reports on the steering dynamics of physically realizable $\mathcal{PT}$-symmetric optical systems are restricted to the unbroken $\mathcal{PT}$-symmetric regimes, as the dynamics in the broken $\mathcal{PT}$-symmetric regime tends to cause instabilities in the functionality of the device \cite{govindarajan2018tailoring,komissarova2019pt}.  The phenomena of OB and OM seem to be unexplored so far in the literature in the broken $\mathcal{PT}$-symmetric phase in the presence of nonuniformities of the grating. Induced by this fact, we look into the input vs output characteristics of the proposed system at $g = 5$ in the absence of detuning parameter ($\delta = 0$).  It is apparent from Fig. \ref{fig3}(a) that the device remains stable even in the broken $\mathcal{PT}$-symmetric systems and the role of chirping is almost the same as in the case of unbroken $\mathcal{PT}$-symmetric regime with a cubic nonlinearity observed in Fig. \ref{fig_1}(a) with the main difference that the slope of the OB curves is now almost unity and thus we obtain ramp-like stable states in the broken $\mathcal{PT}$-symmetric regime. Previously, it was reported that physical systems involving plexcitons \cite{naseri2018optical}, graphene \cite{sharif2016experimental,naseri2018terahertz}, and plasmons \cite{daneshfar2017switching,dai2015low} are the only known structures whose input-output characteristics show ramp like stable states. Recently, we have demonstrated that a broken $\mathcal{PT}$-symmetric uniform FBG can exhibit such unique stable states as a result of delicate balance between the gain and loss \cite{govindarjan2019up,vigneshraja2019}. From our numerical investigations we conclude here that such unique states can exist even in a nonuniform FBG.  In Fig. \ref{fig4}(a) we observe that the system remains in the first stable state till $P_0 = 1.75$ in the absence of chirping. When $C = 0.5$ and $1$, this intensity value gets dropped. However, the intensity builds up in respect to the operation at $C = -0.5$ and $-1$. More importantly, the output characteristics of the system is bistable even though the transmission characteristics in Fig. \ref{fig3}(b) is different from those ones obtained in the unbroken $\mathcal{PT}$-symmetric regime (cf. Figs. \ref{fig_1} and \ref{fig3}).  
The minimum value  of transmission is close to unity for all input powers. The peak of the transmission curves shifts to higher and lower input intensities in the presence of positive and negative chirping, respectively. The first stable branch resembles a ramp for a span of input powers. This span is altered by the interplay between the chirping (C) and detuning parameter ($\delta$) as seen in Figs. \ref{fig3}(c) and \ref{fig3}(d).  When both the parameters have negative values, the span is large (brown dotted). This means that the switching to the second stable state occurs at large input intensities. However, this span decreases, when the chirping is positive and still the device is operated in the negative detuning regime (green solid). Further, the span shrinks for a combination of positive detuning and negative chirping (red solid). Of all the four curves in the middle panel, the plot which is simulated at a positive value of detuning and chirping (black dotted)  shows ramp behavior for the least span of input power and it features a narrow hysteresis width. The structure supports ramp like bistable states on both longer and shorter wavelength sides of the Bragg wavelength when $C = 0.5$ as shown in Fig. \ref{fig3}(e). We obtain a broad bistable region on the shorter wavelength side and as we move towards the longer wavelengths, this region gets  narrowed. However, if the chirping is negative ($C = -2$), the system supports more number of desirable bistable states on the longer side of the Bragg wavelength as seen in Fig. \ref{fig3}(f) compared to Fig. \ref{fig3}(e). It is worthwhile to mention that as pointed out by Maywar \emph{et al., } any OB/OM configuration which exhibits a large spectral range and featuring considerable amount of separation between the switching intensities of neighboring wavelengths can be employed in the construction of all-optical memories \cite{maywar1997transfer}. 

The phenomenon of optical multistability in the presence of quintic nonlinearity remarkably persists in the broken $\mathcal{PT}$-symmetric regime too. The intensities required to switch between any two successive stable branches at $\delta = 0$ gets intensified or trimmed down corresponding to the positive and negative signs of chirping parameter (C), respectively as noticed in Fig. \ref{fig4}(a). This is because of the shift in the transmission peak towards lower or higher input intensities with respect to the variations in the chirping parameter from negative to positive values as seen in Fig. \ref{fig4}(b). Moreover, the transmission of a chirped $\mathcal{PT}$-symmetric FBG is greater than unity for all values of $P_0$ even in the presence of higher order nonlinearities. Due to the presence of self-defocusing quintic nonlinearity, the multistable states can switch back to their previous states or switch to the next state at low intensities, if both the chirping and detuning parameters are negative (green dot-dashed) as seen in Fig. \ref{fig4}(c). If the chirping is positive and detuning is negative, there is an increase in the span of intensities over which ramp-like stable states are observed (red solid). When the detuning parameter takes positive values, there is a huge growth in the switch-down intensity between the second and first stable branches for both positive and negative chirping values as shown in Fig. \ref{fig4}(d). The system possesses optical tristability for a narrow range of input power as seen in Fig. \ref{fig4}(e). The same system exhibits an additional optical quadrastable state for a narrow range of input power when operated in the positive detuning regime. In particular, the system behavior is the same, in terms of the number of stable states in the positive and negative detuning regimes for both the Figs. \ref{fig4}(e) and \ref{fig4}(f) with the difference being that the range of detuning parameters over which the system demonstrates the above mentioned behavior is broadened in the negative detuning regime and gets narrowed in the positive detuning regime. For convenience, these behaviours are tabulated in Table \ref{tab5}, which further illustrates different stable regions exhibited by the system under the variation in the detuning parameter ($\delta$) at fixed values of chirping ($C=-1$ and $1$).
\begin{table}[t]
	\caption{\centering{Number of stable states vs $P_0$ in the broken $\mathcal{PT}$-symmetric regime as shown in Figs. \ref{fig4}(e) and \ref{fig4}(f).}}
	\begin{center}
		\begin{tabular}{|c|c|c|}
			\hline
			{Number of}& {Region 1}& {Region 2} \\
			{stable states}&{}&{} \\
			\cline{2-3} 
			\hline
			{2}&$P_{down}^1$ $\le$ $P_0$ $<$ $P_{down}^2$& $P_{up}^2$ $<$ $P_0$ $\le$ $P_{up}^3$   \\
			\cline{2-3}
			{}&\multicolumn{2}{|c|}{$P_{up}^1$ $<$ $P_0$ $<$ $P_{down}^2$ (only for Fig. \ref{fig4}(f))}\\
			\hline
			{3}&$P_{down}^2$ $\le$ $P_0$ $<$ $P_{down}^3$& $P_{up}^1$ $<$ $P_0$ $\le$ $P_{up}^2$   \\
			\hline
			\cline{2-3}
			{4}& \multicolumn{2}{|c|}{$P_{down}^3$ $<$ $P_0$ $\le$ $P_{up}^1$} \\
			\hline
		\end{tabular}
		\label{tab5}
	\end{center}
\end{table}

\subsection{Right incidence: Unbroken $\mathcal{PT}$-symmetric regime}
\begin{figure}[t]
	\centering
	\includegraphics[width=0.5\linewidth]{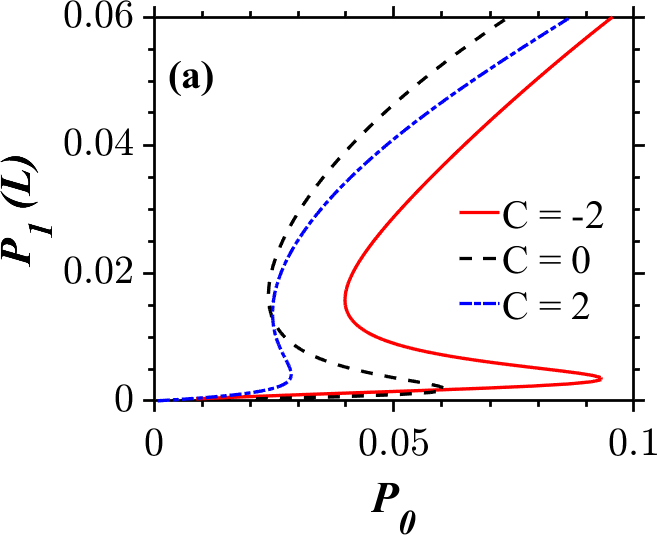}\includegraphics[width=0.5\linewidth]{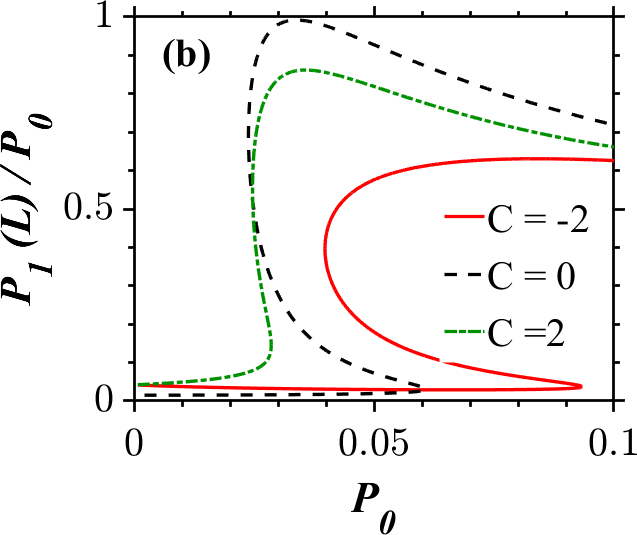}\\\includegraphics[width=0.5\linewidth]{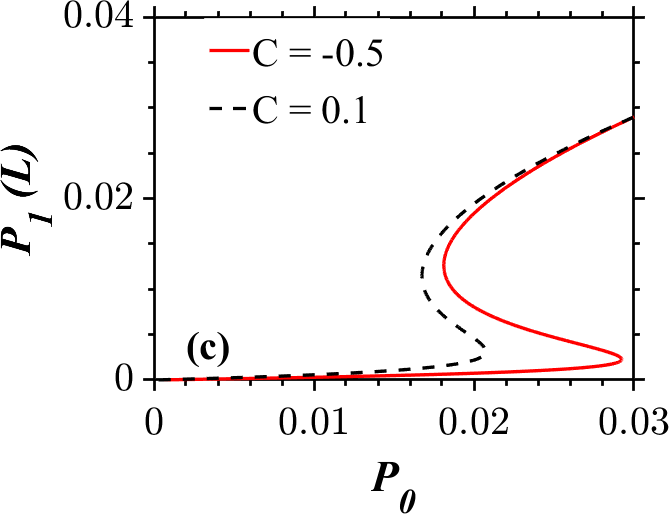}\includegraphics[width=0.5\linewidth]{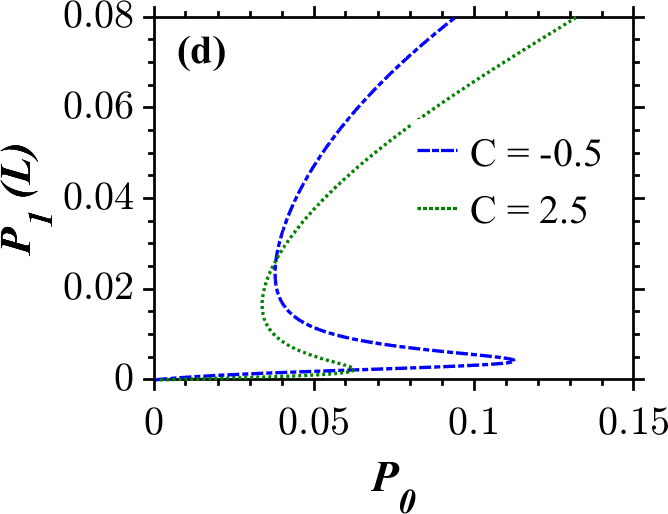}\\\includegraphics[width=0.5\linewidth]{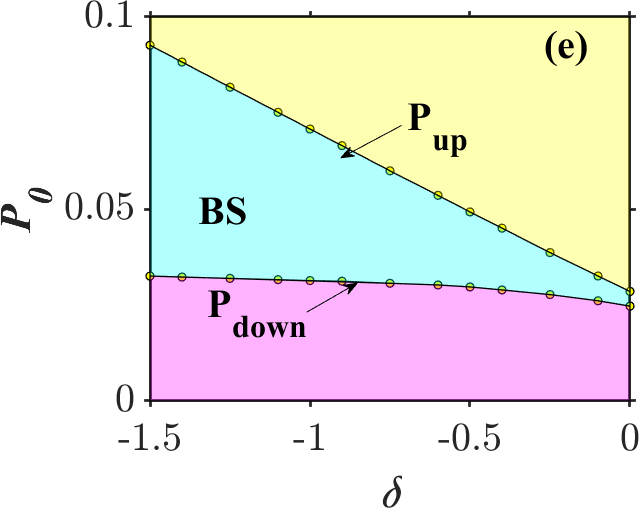}\includegraphics[width=0.5\linewidth]{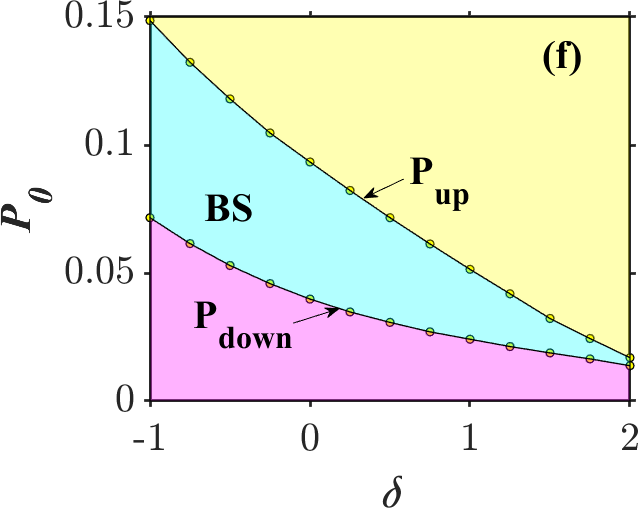}
	\caption{Plots illustrating the same dynamics with the same system parameters as used in Fig. \ref{fig_1} except the light is launched from the right surface of the FBG. Also, the value of detuning parameter is set to $\delta = -1$  in plot (d). In plots (e) and (f), the regions shaded in cyan indicates the range of $P_0$ for which the system shows the bistable behavior and the other regions in magenta and yellow represent $P_0 < P_{down}$ and $P_0 > P_{up}$, respectively.}
	\label{fig5}    
\end{figure}

\begin{figure}[t]
	\centering
	\includegraphics[width=0.5\linewidth]{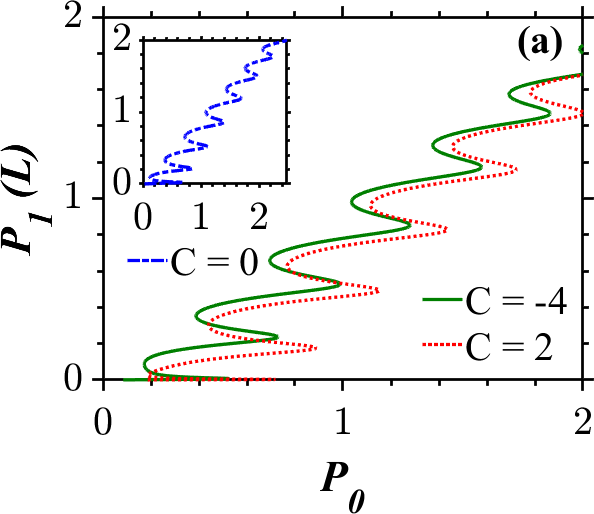}\includegraphics[width=0.5\linewidth]{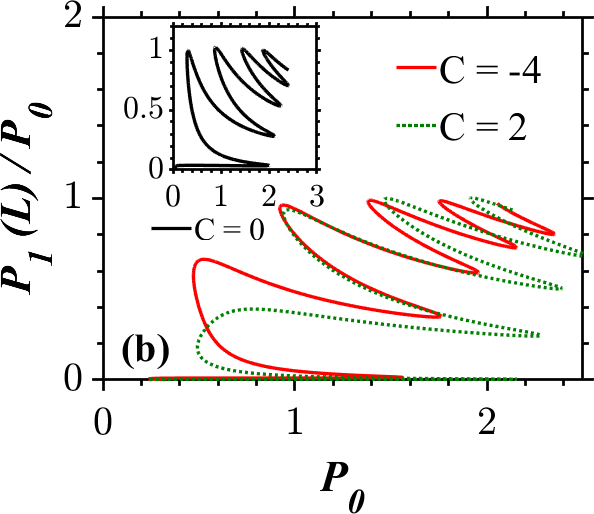}\\\includegraphics[width=0.5\linewidth]{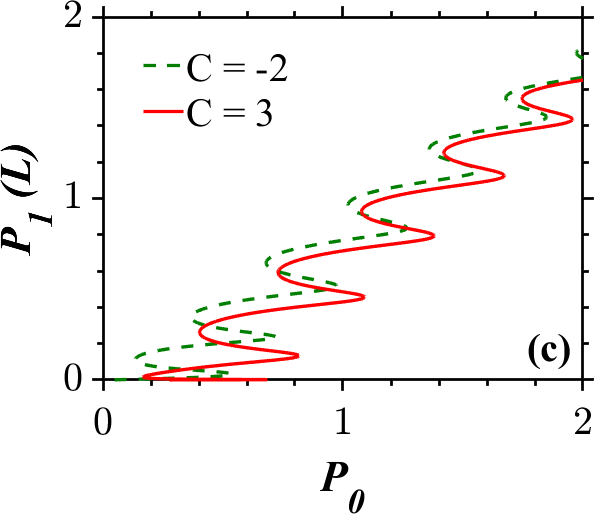}\includegraphics[width=0.5\linewidth]{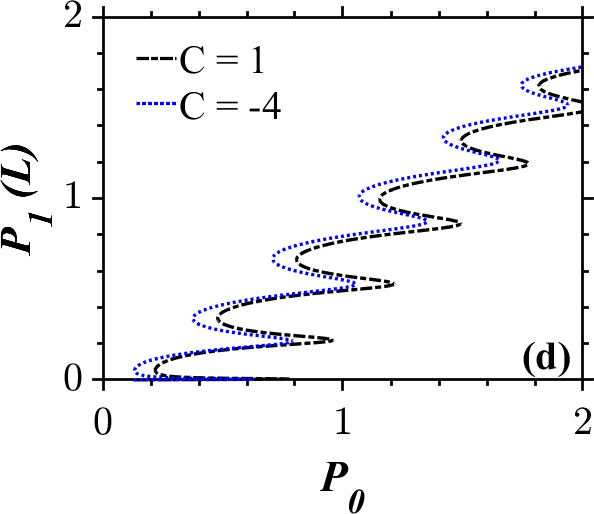}\\\includegraphics[width=0.5\linewidth]{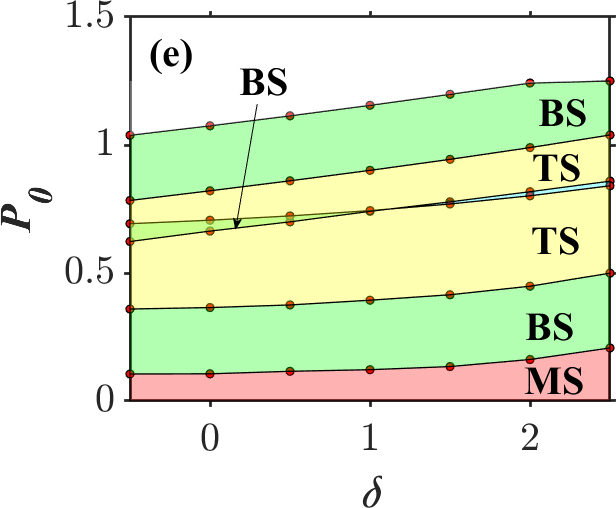}\includegraphics[width=0.5\linewidth]{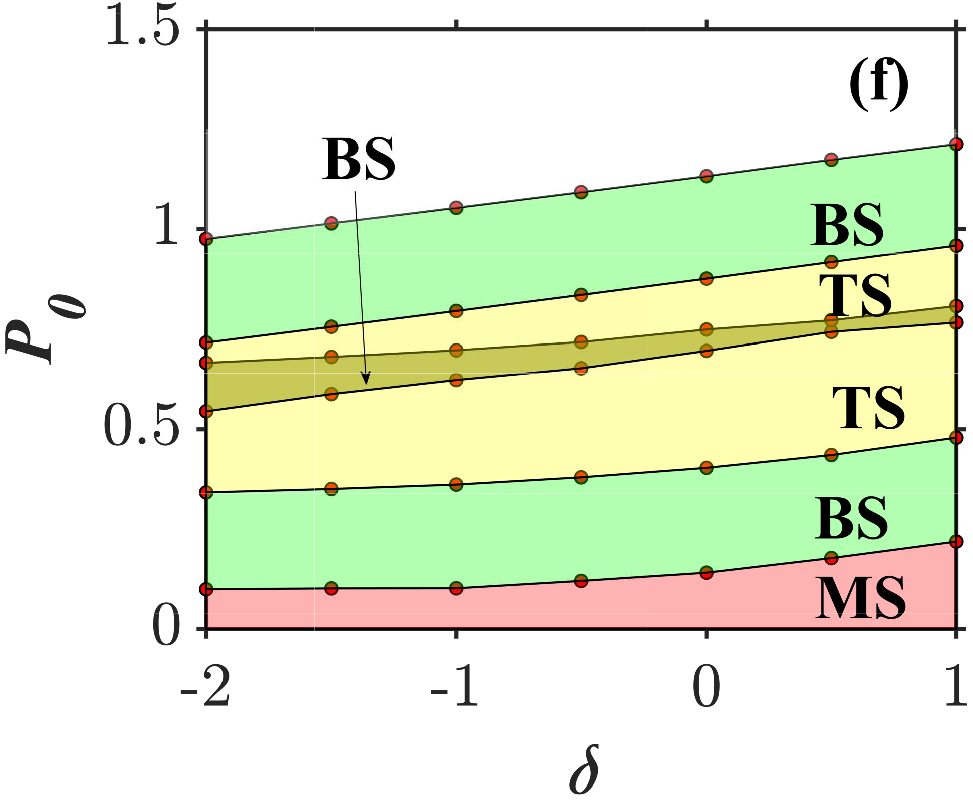}
	\caption{The multistable dynamics of the chirped $\mathcal{PT}$-symmetric FBG. The system parameters are same as seen in Fig. \ref{fig2} except the light is launched from the right input surface of grating structures. In plots, (e) and (f), the range of input intensities ($P_0$) for which the system exhibits single stable state is shaded in red. Also, in the regions shaded in green and yellow the system shows two and three stable states, respectively for any value of given input $P_0$.}
	\label{fig6}    
\end{figure}

The phase-shifted gratings are deemed to be the finest FBG structures to construct all-optical switches with low switching intensities \cite{radic1994optical,radic1995theory}. Even a $\mathcal{PT}$-symmetric FBG can reveal such a  switching at low intensities, thanks to the concept of reverse light incidence on the FBG's input surface. Such a unique physical process is not at all viable in the context of traditional FBG structures, since they produce the same bistable and multistable attributes irrespective of the change in the direction of light incidence. But in the framework of $\mathcal{PT}$-symmetric FBG structures,  switching dynamics for left and right incidence can be distinctive and therefore one can term this switching behavior as \emph{directional dependent switching or nonreciprocal switching}. With this concise introduction on the nonreciprocal switching, we directly look into the switching characteristics for the right light incidence in such chirped $\mathcal{PT}$-symmetric FBG.

From Fig. \ref{fig5}(a), we report that the switching intensities can be remarkably reduced to an order of  \emph{25 times} (approximately) simply by altering the direction of incidence with no changes to the system parameters used in Fig. \ref{fig_1}(a). The switch-up intensities for different chirping values ranging from $C = -2$ to $C = 2$ (in steps of two) are measured to be 0.09, 0.06, and 0.02, respectively. The corresponding widths of the hysteresis curves are also reduced with an increase in the chirping parameter (C).  An important question which arises here is how a simple change in direction of incidence could result in reduction of switching intensity? To address this query, one must recall the thumb rule which states that one must opt for a material with large value of nonlinear coefficient or one must devise structures that can support local field enhancement in the nonlinear regime  to realize low power optical switches \cite{el2018non,dai2015low}. Before the advent of $\mathcal{PT}$-symmetric notion, it was assumed that any active plasmonic device with noble metals like gold, silver in its structure supports local field enhancement in the visible and near-infrared region. The same can be achieved with the help of excitation of graphene surface plasmons at resonance. Now coming to the $\mathcal{PT}$-symmetric systems, the maxima of the total optical field are stationed in the gain regions when the direction of incidence is reversed form left to right on its input surfaces. As a direct consequence, it is obvious that  the propagating fields are amplified. However, in the conventional light incidence direction (left) the field maxima are located within the loss regions due to the asymmetric nature of the device which suppresses any sort of constructive interaction between the forward and backward waves and thereby resulting in an adverse effect, namely  reduction of reflected and transmitted intensities. Hence the switching intensity is large compared to the right light incidence direction. It is worthwhile to remember that this concept of local field enhancement in nonreciprocal Bragg structures was formulated by Kulishov \emph{ et al.} in the linear regime a long back \cite{kulishov2005nonreciprocal}. The strong dependence of steering dynamics on the direction was discovered very recently \cite{komissarova2019pt} but then the impact on reduction of switching intensities was not reported. We affirm that this is the first instance where such a novel way of controlling light with light is proposed in a nonlinear inhomogeneous medium.  In Fig. \ref{fig5}(b), we observe that the transmission is unity in the absence of chirping. For both positive and negative signs of chirping the peak of transmission is reduced from unity to lower values. For $C = -2$, there is no sharp transition from the peak to low values, although the switching threshold gets decreased. Instead, the transmission curve is more or less flat post the occurrence of peak transmission. These studies confirm the fact that the active plasmonics and $\mathcal{PT}$-optical systems share a common ground \cite{el2018non} which can help in devising next generation light-wave systems including low power all-optical switches.  Figure  \ref{fig5}(c) interprets the role of chirping for two different values of chirping parameter at $\delta = 1$. The system shows signs of switching at very low intensities of the \emph{order of 0.02}, if the sign of chirping is positive and the operating wavelength lies on the shorter wavelength side of the Bragg wavelength (black dotted) which is yet another remarkable outcome of our investigation.  These input intensities increase when the detuning and chirping parameters take the alternative signs (red solid in Fig. \ref{fig5}(c) and green solid in Fig. \ref{fig5}(d)), while the switching intensities are comparatively larger for the curve simulated with the negative chirping and the operating wavelength falls on the longer wavelength side of the Bragg wavelength (brown dotted in Fig. \ref{fig5}(d)). But this  intensity is extremely low when compared to the curve plotted with the same system parameters for left incidence in Fig. \ref{fig_1}(a). In addition to the ultra low power switching, the right incidence also alters the span of wavelength over which desirable bistable states are obtained. When the chirping is positive ($C = 2$), the system favors the formation of bistable states only on the shorter wavelength side of Bragg wavelength as shown in Fig. \ref{fig5}(e). Nevertheless, the system supports desirable bistable states on the both sides of Bragg wavelength when the chirping is negative ($C = -2$) as seen in Fig. \ref{fig5}(f). Thus one can conclude that the spectral range can be easily tuned by altering the chirping parameter ($C$). 
\begin{table}[htbp]
	\caption{\centering{Number of stable states vs $P_0$ in Figs \ref{fig6}(e) and \ref{fig6}(f).}}
	\begin{center}
		\begin{tabular}{|c|c|c|}
			\hline
			{Number of}& {Region 1}& {Region 2} \\
			{stable states}&{}&{} \\
			\cline{2-3} 	
			\hline
			\hline
			{2}&$P_{down}^1$ $\le$ $P_0$ $<$ $P_{down}^2$& $P_{up}^2$ $<$ $P_0$ $\le$ $P_{up}^3$ \\
			\cline{2-3}
		{}&\multicolumn{2}{|c|}{$P_{up}^1$ $\le$ $P_0$ $\le$ $P_{down}^3$}\\
	\hline
			{3}&{$P_{down}^2$ $\le$ $P_0$ $<$ $P_{up}^1$}&{$P_{down}^3$ $\le$ $P_0$ $<$ $P_{up}^2$} \\
			\hline
		\end{tabular}
		\label{tab7}
	\end{center}
\end{table}

The presence of quintic nonlinearity can provide low intensity multistable states. Such low intensity multistable states can be employed in multilevel signal processing applications \cite{yousefi2015all}. The switching intensities for the multistable states shown in Fig. \ref{fig6}(a) are still low compared to Fig. \ref{fig2}(a). For the right light incidence, the system exhibits three stable states when the input power $P_0$ is tuned from zero to unity, whereas the device with the same system parameters shows no signs of formation of stable states for $P_0$ less than unity for the left incidence (cf. Figs. \ref{fig2} and \ref{fig6}).  The switching intensities vary marginally among the curves plotted in Fig. \ref{fig6}(a). The first transmission peak is less than unity for all the chirping values as shown in Fig. \ref{fig6}(b) and the transmission at the successive peaks is  enhanced. For a positive chirping value of $C = 2$ and zero detuning ($\delta = 0$), the transmission  reduces drastically. The role of detuning parameter in a chirped $\mathcal{PT}$-symmetric FBG with higher order nonlinearity remains the same for both left and right incidences.  The key difference between the curves in the middle panel of Fig. \ref{fig2} and Fig. \ref{fig6} is that the switching intensities in the former plots are  larger when compared to the latter for all the four sign combinations between the chirping and detuning. Also, the system supports more number of stable states in the unbroken $\mathcal{PT}$-symmetric regime for right light incidence than the left light incidence at a given input power (say $P_0 = 2$). The domain of threshold intensities against the continuous variation of detuning parameter, which is plotted in Figs. \ref{fig6}(e) and \ref{fig6}(f), implies that the system shows a tristable state for very low input powers ($P_0<2$) and for input powers less than unity, we observe different stable regions ranging from monostable to tristable via a bistable region. Moreover, between the two tristable regions,  we observe a narrow bistable region in Fig. \ref{fig6}(e) which is comparatively broadened in Fig. \ref{fig6}(f). The number of stable states demonstrated by the system for different values of $P_0$ shown in Figs. \ref{fig6}(e) and \ref{fig6}(f) is also delineated in Table \ref{tab7}.

\subsection{Right incidence: Broken $\mathcal{PT}$-symmetric regime}
\begin{figure}[t]
	\centering
	\includegraphics[width=0.5\linewidth]{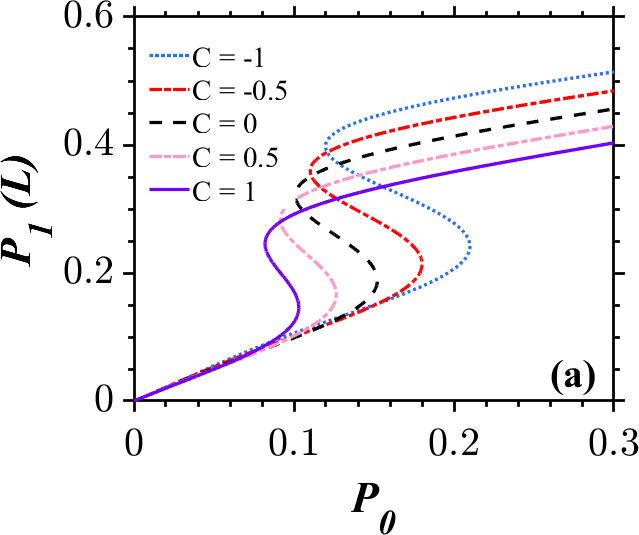}\includegraphics[width=0.5\linewidth]{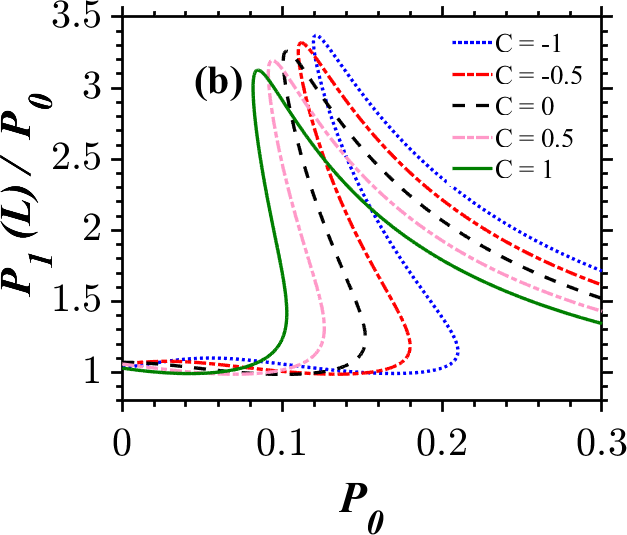}\\\includegraphics[width=0.5\linewidth]{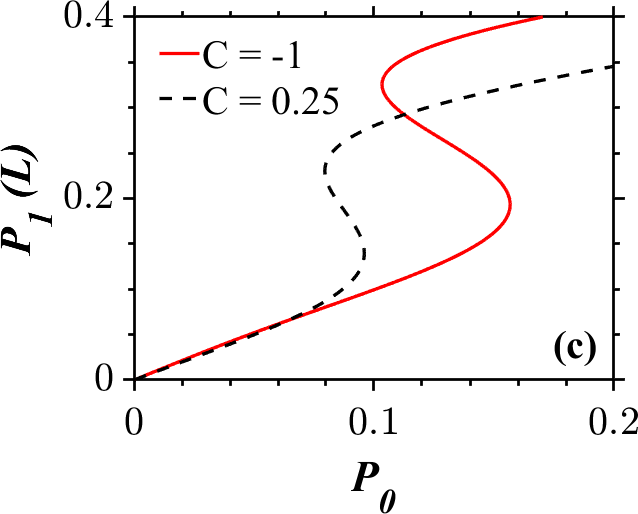}\includegraphics[width=0.5\linewidth]{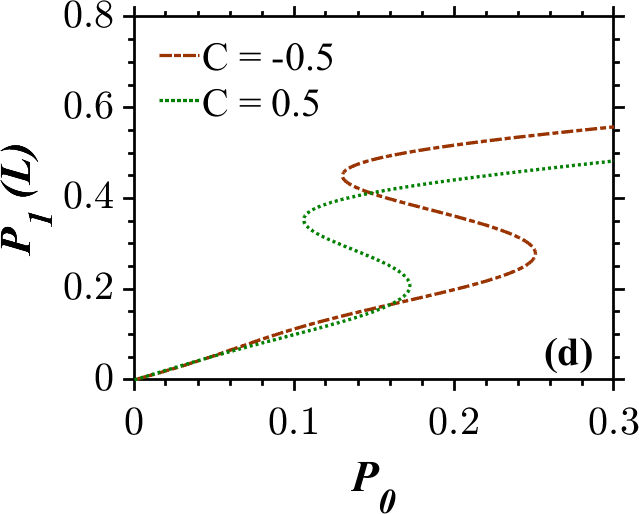}\\\includegraphics[width=0.5\linewidth]{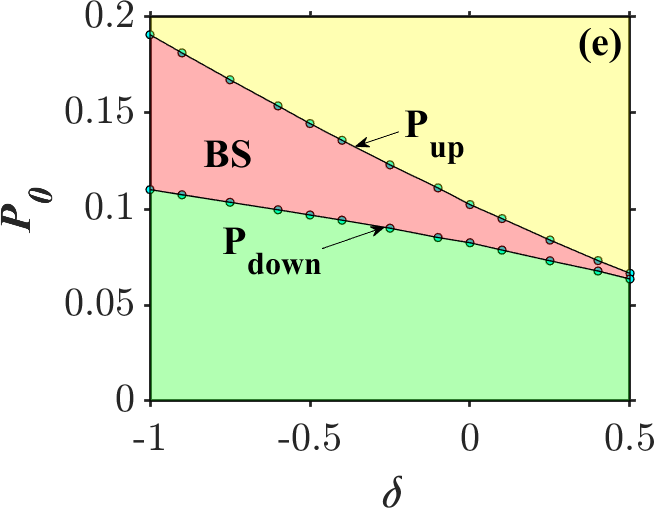}\includegraphics[width=0.5\linewidth]{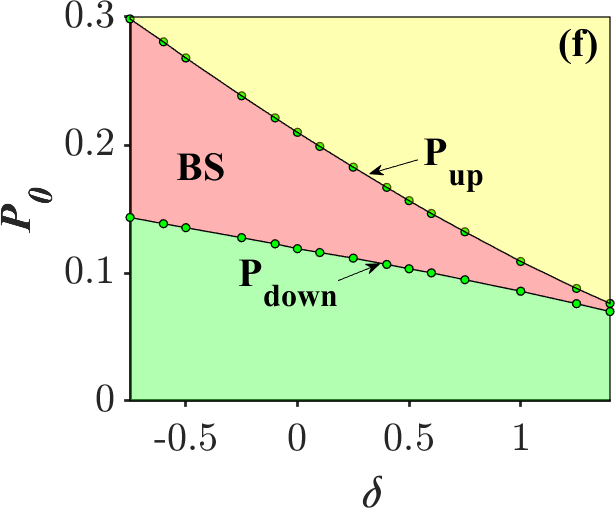}
	\caption{Portraits of the dynamics of the system as in Fig. \ref{fig3} for right light incidence provided that the other system parameters are unchanged except $C = 1$ and $-1$ in plots (e) and (f), respectively. In plots (e) and (f), the regions shaded in red indicates the range of input intensities ($P_0$) for which the system has two stable states for a given input and in other shaded regions of green and yellow the system has $P_0 < P_{down}$ and $P_0 > P_{up}$, respectively. }
	\label{fig7}    
\end{figure}

\begin{figure}[t]
	\centering
	\includegraphics[width=0.5\linewidth]{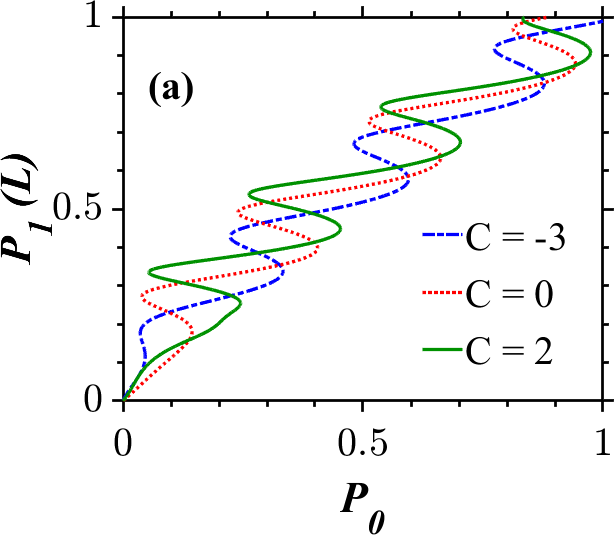}\includegraphics[width=0.5\linewidth]{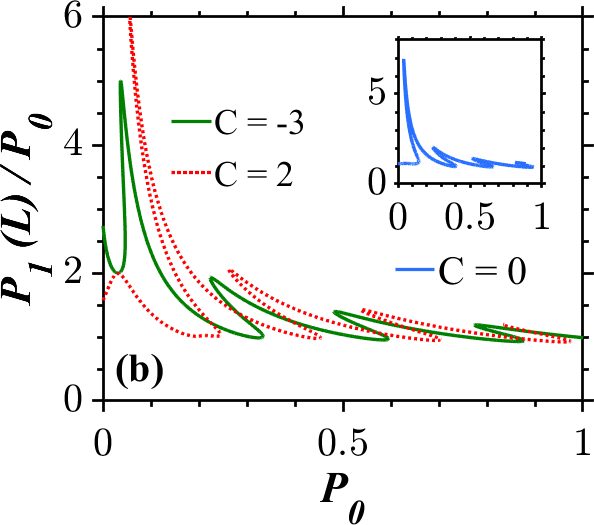}\\\includegraphics[width=0.5\linewidth]{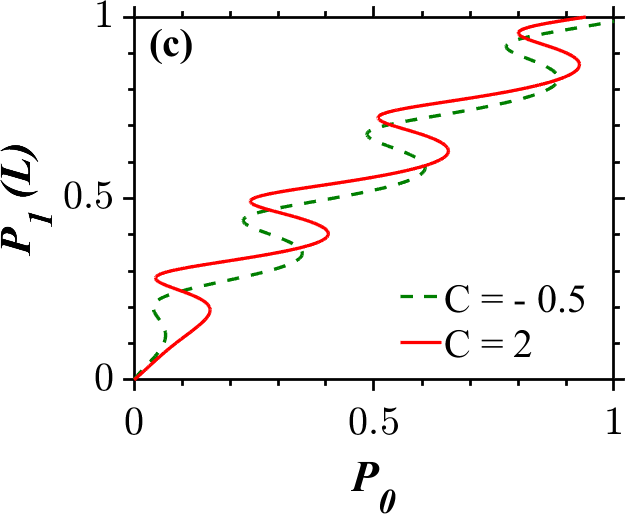}\includegraphics[width=0.5\linewidth]{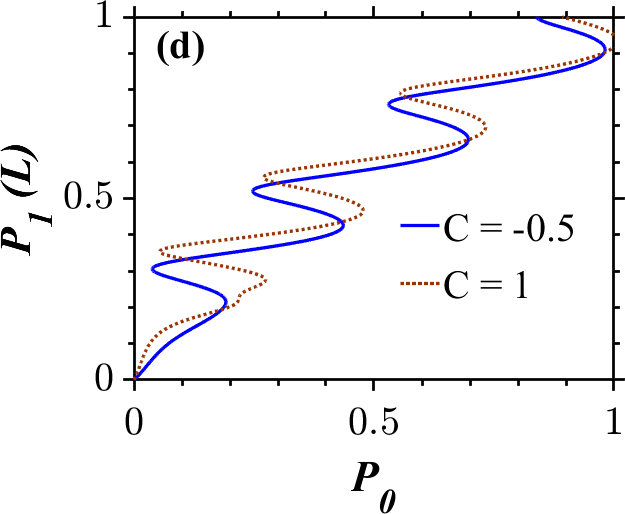}\\\includegraphics[width=0.5\linewidth]{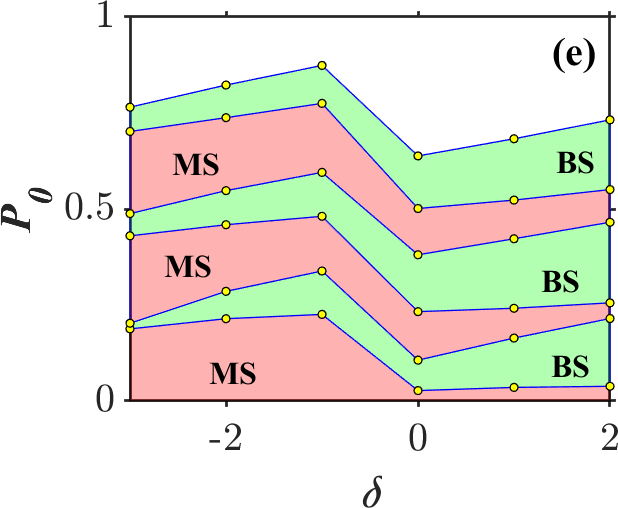}\includegraphics[width=0.5\linewidth]{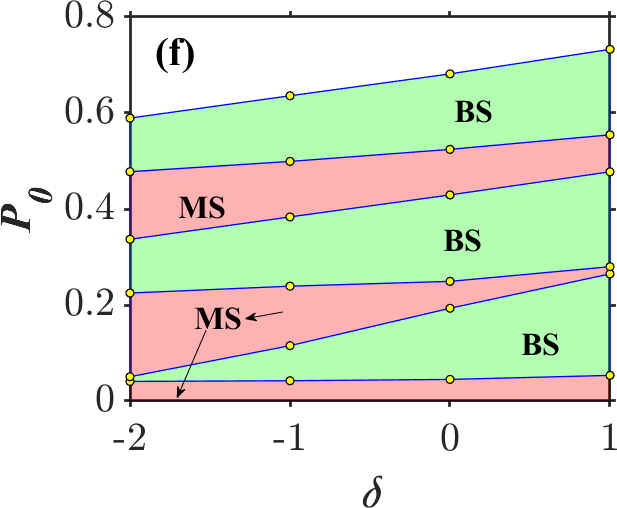}
	\caption{Plots  delineating the same dynamics as in Fig. \ref{fig4} for the right light incidence with the other system parameters unchanged. In plots (e) and (f), the region shaded in green and red indicate the system having two and one stable stable states, respectively, for a given value of input intensity $P_0$.}
	\label{fig8}    
\end{figure}

It is noteworthy to mention that the low intensity switching phenomenon for the right incidence can happen even in the broken $\mathcal{PT}$-symmetric regime of a chirped FBG. To illustrate such a novel behavior, we first consider a simple case where the higher order (quintic) nonlinearity is not present. To understand the role of chirping in the above system, the detuning parameter is set to be zero ($\delta = 0$) as seen in Fig. \ref{fig7}(a). As a result, we obtain novel bistable curves characterized by low switching intensities and ramp like first stable state. The span of input powers for the ramp-like first stable state increases if the chirping is negative and vice-versa. Comparing Figs. \ref{fig3}(b) and \ref{fig7}(b), we conclude that the amplitude of transmission gets enhanced with a change in the direction of incidence for the same system parameters used in Fig. \ref{fig3}. Moreover, the input intensity at which the peak of the transmission occurs is very low and it gets shifted with a change in the chirping value from $C = -1$ to $C = 1$ in steps of 0.5 (see Fig. \ref{fig7}(b)). Similar to the conclusions drawn from Figs. \ref{fig3}(c) and \ref{fig3}(d), the interplay between the chirping and detuning parameter enhances or reduces the span of input intensities over which the curves show ramp like stable states (see Figs. \ref{fig7}(c) and \ref{fig7}(d)).  When the chirping parameter is fixed and if the detuning parameter is varied as in the case of Fig. \ref{fig7}(e) and \ref{fig7}(f), we observe that the bistable region is broad for shorter wavelengths and narrow for longer wavelengths. The key difference between the two is that the later supports more bistable states for positive detuning values than the former. 

Finally, we focus on the OM phenomenon in a broken $\mathcal{PT}$-symmetric FBG in conjunction with higher order quintic nonlinearity for the right light incidence. The system exhibits more than three stable states for input intensity less than unity as shown in Fig. \ref{fig8}(a), whereas the same system for the left light incidence has only one stable state as seen in Fig. \ref{fig4}(a). The number of transmission peaks as well as intensity at the transmission peaks is significantly enhanced in Fig. \ref{fig8}(b) compared to Fig. \ref{fig5}(b). The variations in the OM curves brought by different combinations of detuning and chirping parameters (see Figs. \ref{fig8}(c) and \ref{fig8}(d)) reveal that the combined role of chirping and detuning remains unchanged with the direction of incidence but it depends only on the type of nonlinearity (either cubic or quintic). But the key role of changing the direction of incidence is that the switching intensities are always low for the right light incidence compared to the left light incidence even in the broken $\mathcal{PT}$-symmetric regime. One can observe from Figs. \ref{fig8}(e) and \ref{fig8}(f) that the system shows OB phenomenon only if the  input intensity falls in the range $P_{down}^n$ $\le$ $P_0$ $\le$ $P_{up}^n$, where $n=1,2,3...$ indicating the order of the stable branches. If the input power doesn't fall within these regions, then there is no formation of bistable states.    
\begin{table}[htbp]
	\caption{Comparison of switching intensities for different combinations of chirping ($C$) and detuning parameter ($\delta$) in the presence of cubic nonlinearity ($\gamma = 1$, $\Gamma = 0$) and cubic-quintic nonlinearities ($\gamma = \Gamma = 1$).}
	\begin{center}
		\begin{tabular}{|c|c|c|c|}
				\hline
			{Type of}&\multicolumn{3}{|c|}{Switching intensities} \\
			 
		{nonlinearity}&\multicolumn{3}{|c|}{\textbf{($P_{up}$ / $P_{down}$)}} \\
			\cline{2-4} 
			\textbf{} & {}& {$C>0$}&{$C<0$} \\
			\hline
			\textbf{Cubic}& {$\delta>0$}&\textbf{\textit{lowest}}& low/high   \\
		\cline{2-4}
			{}& {$\delta<0$}&high / low& \textbf{\textit{highest}}   \\
			\hline
			\textbf{Quintic}& {$\delta>0$}&\textbf{\textit{highest}}& high/low   \\
		\cline{2-4}
			{}& {$\delta<0$}&low/high& \textbf{\textit{lowest }} \\
			\hline
		\end{tabular}
		\label{tab1}
	\end{center}
\end{table}
Table \ref{tab1}  summarizes the essence of the role of chirping and detuning parameters in different nonlinear regimes as deduced from our studies from Figs. \ref{fig_1} to \ref{fig8}. Irrespective of the light incidence and the $\mathcal{PT}$-symmetric regimes in which the system is operated, the sign of chirping and detuning must be the same and match with the type of nonlinearity to feature lowest switching intensities. For self-focusing cubic  nonlinearity ($n_2>0$) both $C$ and $\delta$ needs to be positive, while they must be negative in the presence of self-defocusing quintic nonlinearity ($n_4<0$) to realize low power all-optical switches. If there exists any mismatch between the signs of $C$ and $\delta$, the value of input intensities can be higher and it depends on the magnitude of these parameters. Lastly, if both $C$ and $\delta$ have the same signs but there exists a mismatch with the type of nonlinearities, then the system exhibits highest switching intensities as well as the broadest hysteresis width. Note that such curves are preferable for all-optical memory applications rather than switching.  
\section{Exceptional point dynamics of a chirped $\mathcal{PT}$-symmetric FBG}
\label{Sec:4}
\begin{figure}[t]
	\centering
	\includegraphics[width=0.5\linewidth]{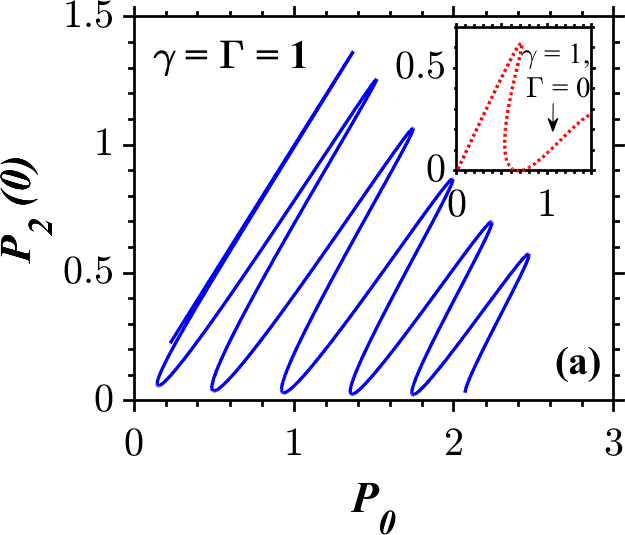}\includegraphics[width=0.5\linewidth]{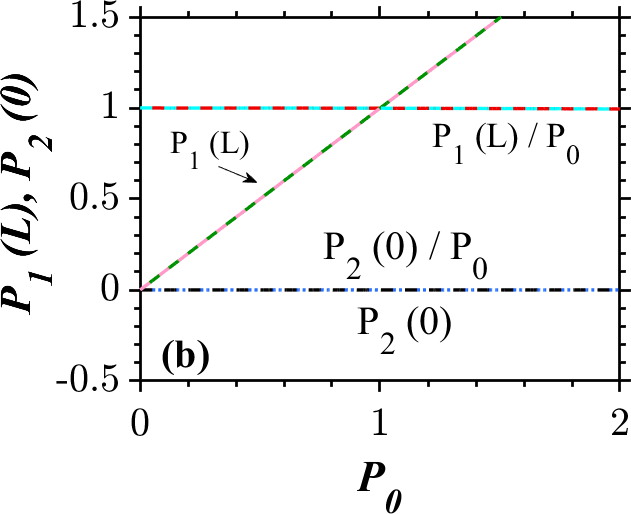}\\\includegraphics[width=0.5\linewidth]{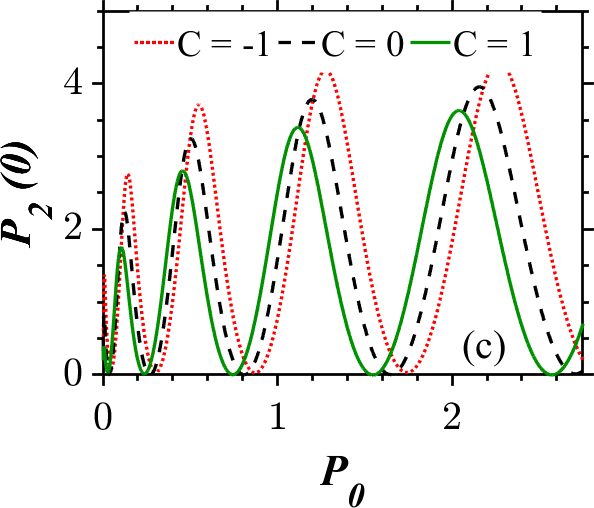}\includegraphics[width=0.5\linewidth]{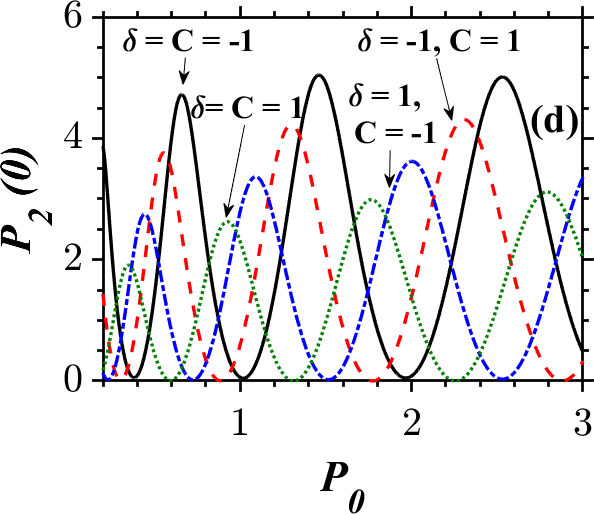}\\\includegraphics[width=0.5\linewidth]{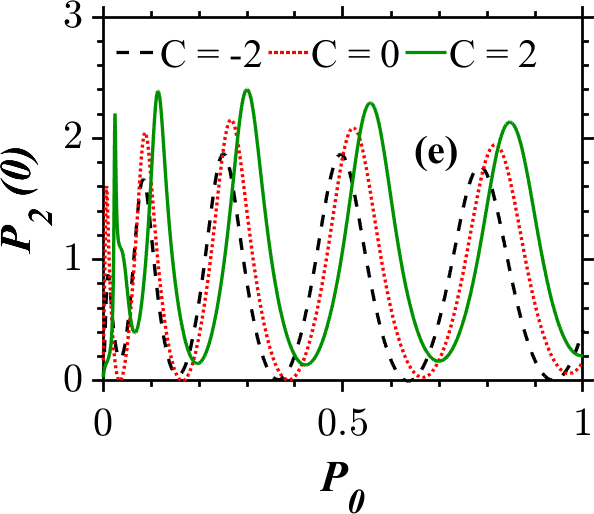}\includegraphics[width=0.5\linewidth]{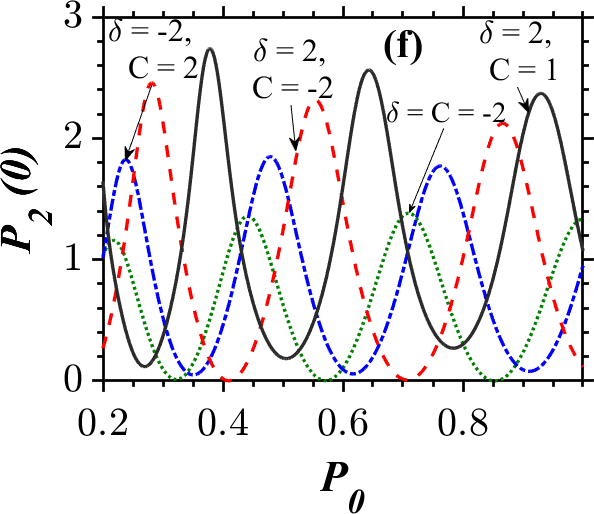}
	\caption{Plot (a) depicts the reflective OB and OM curves in  a conventional FBG ($g = 0$) for different nonlinear regimes. Plots (b)  illustrates unidirectional wave transport at the exceptional point ($k = g = 4$) in a chirped $\mathcal{PT}$-symmetric  FBG. Here, the input -- output characteristics ($P_1(L)$ vs $P_0$) for the left and right light incidences are, respectively, indicated by the black dotted and cyan solid lines  and their corresponding transmission ($P_1(L)$ / $P_0$) is  indicated by the red solid and green dotted lines in the presence of chirping ($C$), detuning ($\delta$) parameters, and the nonlinearities (cubic or quintic). The remaining plots (c)--(f) portray the input intensity ($P_0$) against the reflected intensity ($P_2 (0)$) with the system parameters $\delta = \Gamma= 0$, $\gamma = 1$ for the right light incidence, where Plots (c) and (d) depict the periodic light transfer for varying chirping parameter and combined effects of the detuning and chirping parameters with cubic nonlinearity ($\gamma = 1$, $\Gamma = 0$) and  Plots (e) and (f) illustrate the same dynamics as shown in (c) and (d) with cubic and quintic nonlinearities. ($\gamma = \Gamma = 1$).}
	\label{fig9}    
\end{figure}
It is to be remembered that under the mathematical condition $k = g$, the system to said to be operating at the exact $\mathcal{PT}$-symmetric phase. The incident light launched into the FBG experiences unidirectional reflectionless wave transport at the singular point designated to be \textit{unidirectional invisibility} resulting in the ideal transmission of light from one (left) side. However, the reflection for the other (right) direction of light incidence is enhanced in the case of linear gratings with balanced gain and loss \cite{lin2011unidirectional,kulishov2005nonreciprocal,huang2014type}. Lin \emph{et al.} came up with the theoretical demonstration of this peculiar phenomenon in a uniform $\mathcal{PT}$-symmetric nonlinear gratings \cite{lin2011unidirectional} and they concluded that the unidirectional invisibility persist even in the presence of nonlinearities.
Not so long ago, this concept was extended to a nonuniform  (apodized) gratings \cite{lupu2016tailoring}. However, it is bounded to linear regime alone.  All these studies enhance our curiosity to understand whether the nonlinear $\mathcal{PT}$-symmetric FBG will show unidirectional wave transport in the presence of a chirping nonuniformity. So far, we have studied the OB/OM curves pertaining to the transmitted (output) light. Along these lines, we also would like to study the reflective OB/OM offered by the $\mathcal{PT}$-symmetric FBG at the exceptional point to find out the net effect of the chirping and other system parameters on the OB/OM curves for different directions of light incidence. Before we present our results, we define the  mathematical relations for the reflected intensity as $P_2(0) = |\psi_{-}(0)|^2$ and reflection coefficient as $P_2(0)/P_0$ = $|\psi_-(0)/\psi_+(0)|^2$.  

A conventional FBG ($g = 0$) exhibits reflective bistability in the presence of both cubic and quintic nonlinearities as shown in Fig. \ref{fig9}(a). From the numerical results depicted in Fig. \ref{fig9}(b), we reassert that unidirectional invisibility (solid lines) sustains even in the presence of chirping in a nonlinear $\mathcal{PT}$-symmetric FBG  for the left light incidence. Also, from Fig. \ref{fig9}(b) we observe that the reflected intensity is always zero and OB/OM disappears. These conclusions hold true irrespective of the variations in the other system parameters like detuning and nonlinearities (both cubic and quintic). When the direction of incidence of input light is reversed, the transmitted intensity ($P_1 (L)$) and the transmission coefficient ($P_1 (L) / P_0$) remain unchanged from those obtained for the left light incidence. But the reflected intensity ($P_2 (0)$) shows variations with respect to changes in the nonlinearities, detuning parameter, and chirping of the proposed system as shown in Figs. \ref{fig9}(c) -- \ref{fig9}(f). In contrast to Fig. \ref{fig9}(a), the input intensity ($P_0$) vs reflected intensity ($P_2 (0)$) curves do not exhibit OB/OM at the exact $\mathcal{PT}$-symmetric phase. Instead, the intensity of the reflected light shows the periodic transfer of light with the formation of crests and troughs within a small span of input intensity $P_0$ and with further increase in the input power, this span between a crest and trough increases. The positive chirping ($C = 1$) decreases the reflected intensity at the peaks (green solid), whereas the negative chirping ($C = -1$) enhances it in the absence of detuning with the inclusion of cubic nonlinearity as shown in Fig. \ref{fig9}(c). Also, the curves are shifted towards higher and lower input intensities by the negative and positive chirping, respectively. Even in the presence of detuning, the shifting of input intensities of the curves and variations in the peak at output intensities sustain as seen in Fig. \ref{fig9}(d). When both detuning and chirping parameters are positive, the curves feature lowest input as well as output peak intensities (green dotted). On the other hand,  if both these parameters are negative, then the curves exhibit highest input and  output peak intensities (black solid). For the other two combinations ($\delta > 0$, $C < 0$ and $\delta < 0$, $C > 0$) the intensities (red dotted and blue dot-dashed) lie between these two curves.  The output intensity at each peak marginally decreases with an increase in the input power $P_0$ as seen in Fig. \ref{fig9}(e), which is simulated in the presence of cubic--quintic nonlinearities. The interpretation for shifting of the input intensities and variations in the output peak intensities discussed earlier for the Figs. \ref{fig9}(c) and \ref{fig9}(d) does not hold good for Figs. \ref{fig9}(e) and \ref{fig9}(f), respectively. But the exact opposite effect is observed both in the presence and absence of detuning parameter ($\delta$). In other words, the curves which featured highest input and output intensities in Figs. \ref{fig9}(c) and \ref{fig9}(d) now exhibit lowest input and output peak intensities Figs. \ref{fig9}(e) and \ref{fig9}(f), respectively and vice-versa.
\section{conclusions}
\label{Sec:5}
In conclusion, we have examined the significance of the Bragg grating nonuniformity in the form of chirping on the OB/OM phenomenon exhibited by a $\mathcal{PT}$-symmetric FBG. The grating nonuniformity plays a significant role in altering the spectral range of the OB/OM curves with respect to the detuning. Since it is possible to fine-tune the operating wavelength of the signal at ease, the present work presents a comprehensive picture of role of detuning parameter alongside the chirping parameter in the reduction of switching intensity in the realization of \emph{low power all-optical switches}. This is a novel way to cut-down the switching intensity, specifically in the presence of $\mathcal{PT}$-symmetry. Also, the chirped $\mathcal{PT}$-symmetric FBG affords design flexibility to build any sort of nonlinear applications such as optical switches, memories or signal regenerators by simply manipulating any one of the system parameters. The number of stable states can be controlled at ease with the aid of nonlinearity offered by the system or by tuning the input power  for a multilevel signal processing applications. The sign mismatch among the nonlinear coefficients, chirping and detuning parameter aids in varying the spectral range of the input-output curves and broadening of the hysteresis width to realize optical memory applications. From the numerical simulations we confirm that the phenomenon of unidirectional invisibility recurs even in the presence of chirping nonuniformity of the grating. Also, we found that that system does not posses reflective OB/OM for the right incidence at the exceptional point unlike the conventional FBG. For those applications which demand large degree of spectral uniformity, the broken $\mathcal{PT}$-symmetry regime is the optimum choice. Further, the formation of ramp-like stable states and the notion of local field enhancement in the broken $\mathcal{PT}$-symmetric regime confirm that the simple nonuniform $\mathcal{PT}$-symmetric Bragg-grating structures are closely associated with active plasmonic structures. More importantly, an ultra low power all-optical switch can be realized by launching the light from the right surface of the $\mathcal{PT}$-symmetric FBG, thereby opening a new roadmap to controlling light with light for all-optical signal processing. 

\section*{Acknowledgments}
SVR is indebted to the financial assistanship provided
by Anna University through Anna Centenary Research Fellowship (CFR/ACRF-2018/AR1/24). The work of AG is supported by the Department of Science and Technology (DST) and Science and Engineering Research Board (SERB), Government of India, through a National Postdoctoral Fellowship (Grant No. PDF/2016/002933).  ML is supported by DST-SERB through a Distinguished Fellowship (Grant No. SB/DF/04/2017.)

\end{document}